%% file: main.tex
\documentclass[%
reprint, 
superscriptaddress,
amsmath,
amssymb,
floatfix,
twocolumn, 
noeprint,
longbibliography
]{revtex4-1}

\usepackage[normalem]{ulem}
\usepackage{graphicx} 
\usepackage{dcolumn}  
\usepackage[english]{babel}
\usepackage[utf8]{inputenc}
\DeclareUnicodeCharacter{2212}{-}

\usepackage{bm} 
\usepackage[]{upgreek}
\usepackage{gensymb}
\usepackage[]{nicefrac}
\usepackage{multirow}
\usepackage{cancel}
\usepackage{color,soul}

\usepackage{hyperref}
\hypersetup{%
    pdfborder = {0 0 0},
    colorlinks=True,
    urlcolor=[rgb]{0,0,0.5},
    linkcolor=[rgb]{0,0,0.5},
    citecolor=[rgb]{0,0,0.5}
}

\begin{document}

\title{Ultralong dephasing times in solid-state spin ensembles via quantum control}


\author{Erik Bauch}
\thanks{these authors contributed equally to this work}
\affiliation{Department	of	Physics,	Harvard	University,	Cambridge,	Massachusetts	02138,	USA}

\author{Connor A. Hart}
\thanks{these authors contributed equally to this work}
\affiliation{Department	of	Physics,	Harvard	University,	Cambridge,	Massachusetts	02138,	USA}

\author{Jennifer M. Schloss}
\affiliation{Department	of	Physics,	Harvard	University,	Cambridge,	Massachusetts	02138,	USA}
\affiliation{Department	of	Physics,	Massachusetts	Institute	of	Technology,	Cambridge,	Massachusetts	02139,	USA}
\affiliation{Center	for	Brain	Science,	Harvard	University,	Cambridge,	Massachusetts	02138, USA}

\author{Matthew J. Turner}
\affiliation{Department	of	Physics,	Harvard	University,	Cambridge,	Massachusetts	02138,	USA}
\affiliation{Center	for	Brain	Science,	Harvard	University,	Cambridge,	Massachusetts	02138, USA}

\author{John F. Barry}
\affiliation{Department	of	Physics,	Harvard	University,	Cambridge,	Massachusetts	02138,	USA}
\affiliation{Lincoln Laboratory, Massachusetts Institute of Technology, Lexington, Massachusetts 02420, USA}

\author{Pauli Kehayias}
\affiliation{Department	of	Physics,	Harvard	University,	Cambridge,	Massachusetts	02138,	USA}
\affiliation{
Harvard-Smithsonian Center	for	Astrophysics,	Cambridge,	Massachusetts	02138,	USA}

\author{Swati Singh}
\affiliation{Williams College,
Department of Physics,
33 Lab Campus Drive,
Williamstown, Massachusetts 01267}

\author{Ronald L. Walsworth}
\email{rwalsworth@cfa.harvard.edu}
\affiliation{Department	of	Physics,	Harvard	University,	Cambridge,	Massachusetts	02138,	USA}
\affiliation{Center	for	Brain	Science,	Harvard	University,	Cambridge,	Massachusetts	02138, USA}
\affiliation{
Harvard-Smithsonian Center	for	Astrophysics,	Cambridge,	Massachusetts	02138,	USA}




\begin{abstract}

Quantum spin dephasing is caused by inhomogeneous coupling to the environment, with resulting limits to the measurement time and precision of spin-based sensors.  The effects of spin dephasing can be especially pernicious for dense ensembles of electronic spins in the solid-state, such as nitrogen-vacancy (NV) color centers in diamond. We report the use of two complementary techniques, spin bath driving, and double quantum coherence magnetometry, to enhance the inhomogeneous spin dephasing time ($T_2^*$) for NV ensembles by more than an order of magnitude. In combination, these quantum control techniques (i) eliminate the effects of the dominant NV spin ensemble dephasing mechanisms, including crystal strain gradients and dipolar interactions with paramagnetic bath spins, and (ii) increase the effective NV gyromagnetic ratio by a factor of two. Applied independently, spin bath driving and double quantum coherence magnetometry elucidate the sources of spin ensemble dephasing over a wide range of NV and bath spin concentrations. These results demonstrate the longest reported $T_2^*$ in a solid-state electronic spin ensemble at room temperature, and outline a path towards NV-diamond DC magnetometers with broadband femtotesla sensitivity.

\end{abstract}

\flushbottom
\maketitle

\section*{Introduction}	

Solid-state electronic spins, including defects in silicon carbide~\cite{Klimov2015,Widmann2015,Heremans2016,Koehl2017,Tarasenko2017}, phosphorus spins in silicon~\cite{Abe2010,Tyryshkin2012}, and silicon-vacancy~\cite{Hepp2014,Heremans2016,Rose2017} and nitrogen-vacancy (NV) centers~\cite{Doherty2013} in diamond, have garnered increasing relevance for quantum science and sensing experiments. In particular, NV centers in diamond have been extensively studied and deployed in diverse applications facilitated by long NV spin coherence times~\cite{Balasubramanian2009,Stanwix2010} at ambient temperature, as well as  optical preparation and readout of NV spin states~\cite{Doherty2013}. Many applications utilize dense NV spin ensembles for high-sensitivity DC magnetic field sensing~\cite{Barry2016, Bucher2017} and wide-field DC magnetic imaging ~\cite{LeSage2013,Glenn2015,Shao2016,Tetienne2016,Glenn2017}, including measurements of single-neuron action potentials~\cite{Barry2016}, paleomagnetism~\cite{Fu2017, Glenn2017}, and current flow in graphene~\cite{Tetienne2016}.



\begin{figure}[ht]
  \centering
  \includegraphics[]{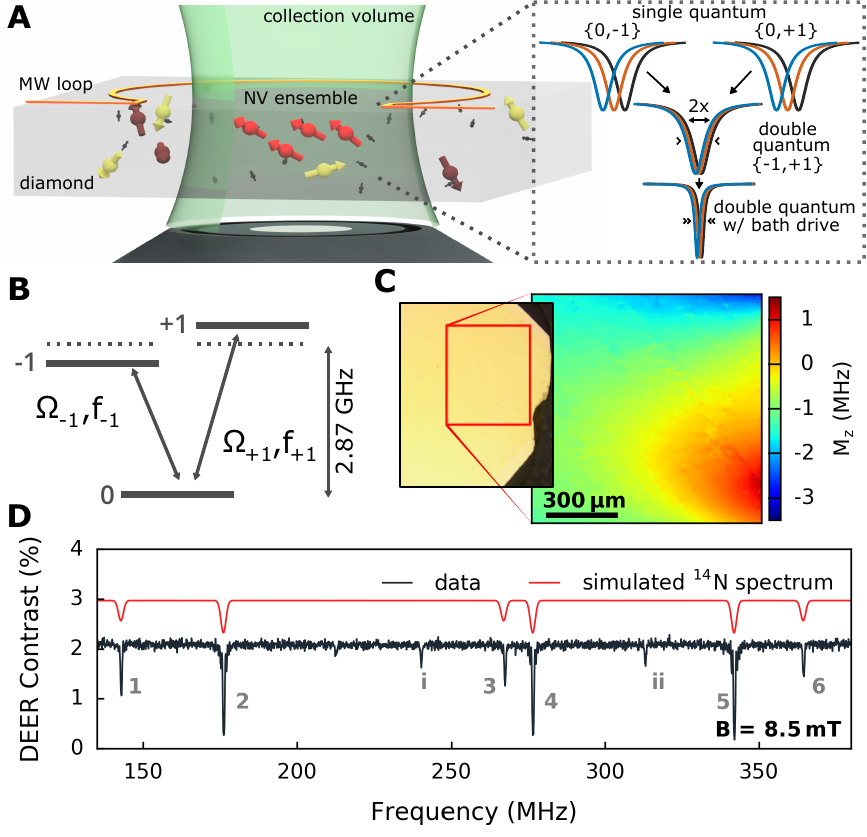}  
  \caption{{NV ensemble spectroscopy of diamond spin bath. (a)} The inhomogeneously broadened electron spin resonance (ESR) linewidth of nitrogen-vacancy (NV) ensembles is a complex function of the local environment within the diamond sample, which includes a diverse bath of electronic and nuclear spins. Inset: Schematics of NV ensemble ESR spectra in the single quantum and double quantum bases, and for double quantum with spin-bath drive. {(b)} Spin-1 ground state of the NV center. {(c)} Imaging of the longitudinal strain component $M_z$ of one NV orientation class across a $1$-$\,$mm$^2$ field of view for Sample B. An optical microscope image of the diamond surface (left) is included for reference with a red box outlining the field of view shown in the NV strain image. {(d)} NV double electron-electron resonance (DEER) spectrum of Sample B, showing six nitrogen groups ($1-6$) attributed to $^{14}$N electronic spins with an external field $B_0 = 8.5\,$mT aligned along a [111]-crystallographic axis (see main text). Linewidths are Fourier-broadened. The peaks labeled $i$ and $ii$  correspond to dipole-forbidden transitions of the $^{14}$N electronic spins ($\Delta m_I \neq 0$, see Suppl.\,XI). 
The simulated spectrum using the full nitrogen Hamiltonian is shown in red, with linewidth and amplitudes chosen to reflect the experimental data.}
\label{fig:fig1}
\end{figure}

For NV ensembles, the DC magnetic field sensitivity is typically limited by dephasing of the NV sensor spins. In such instances, spin interactions with an inhomogeneous environment (see Fig.\,\ref{fig:fig1}a) limit the experimental sensing time to the spin dephasing time~$T_2^{*} \, \lesssim\,1\,\upmu$s~\cite{Acosta2010,Kubo2011,Grezes2015,Choi2017a}. Hahn echo and dynamical decoupling protocols can restore the NV ensemble phase coherence by isolating the NV sensor spins from environmental noise and, in principle, permit sensing times approaching the spin lattice relaxation ($T_1\sim\,$ms)~\cite{DeLange2010,Pham2012,Bar-Gill2013}. However, these protocols restrict sensing to AC signals within a narrow bandwidth. For this reason, the development of high sensitivity, broadband magnetometers requires new approaches to extend $T_2^*$ for NV ensembles while retaining the ability to measure DC signals.

{ 
To date, spin dephasing mechanisms for NV ensembles have not been systematically studied, as spatially inhomogeneous effects do not lead to single NV spin dephasing, which has traditionally been the focus of the NV-diamond literature~\cite{Balasubramanian2009,Maurer2012,Fang2013,Mamin2014}.  Here, we characterize and control the dominant NV spin ensemble dephasing mechanisms by combining two quantum control techniques, 
double quantum (DQ) coherence magnetometry~\cite{Fang2013,Mamin2014} and spin bath driving~\cite{DeLange2012, Knowles2014}. We apply these techniques to three isotopically engineered $^{12}$C samples with widely varying nitrogen and NV concentrations. In combination, we show that these quantum control techniques can extend the NV spin ensemble $T_2^*$ by more than an order of magnitude.
}

{ Several inhomogeneous spectral broadening mechanisms can} contribute to NV spin ensemble dephasing in bulk diamond. { First,} the formation of negatively-charged NV$^-$ centers { (with electronic spin $S=1$)} requires the incorporation of nitrogen into the diamond lattice. As a result, paramagnetic substitutional nitrogen impurities (P1 centers, $S=1/2$)~\cite{Smith1959,Cook1966,Loubser1978} typically persist at densities similar to or exceeding the NV concentration, leading to a `spin bath' that couples to the NV spins via incoherent dipolar interactions{, with a magnitude that can vary significantly across the NV ensemble. Second, $^{13}$C nuclei ($I=1/2$) can be a considerable source of NV spin dephasing in diamonds with natural isotopic abundance ($1.07\,\%$), with the magnitude of this effect varying spatially due to the random location of $^{13}$C within the diamond lattice~\cite{Mizuochi2009,Dreau2012}. Such NV spin ensemble dephasing, however, can be greatly reduced through isotope engineering of the host diamond material~\cite{Balasubramanian2009}. Third, } strain is well-known to affect the diamond crystal and the zero-magnetic-field splitting between NV spin states~\cite{Jamonneau2015,Trusheim2016}. { The exact contribution of strain gradients to NV spin ensemble dephasing has not been quantified rigorously because strain varies throughout and between samples, and is in part dependent upon the substrate used for diamond growth\,\cite{Gaukroger2008,Hoa2014}.} Furthermore, the interrogation of spatially large NV ensembles requires the design of uniform magnetic bias fields to minimize magnetic field gradients across the detection volume. 

{
We assume that the relevant NV spin ensemble dephasing mechanisms are independent and can be summarized by Eqn.\,\ref{eqn:rate}, 
\begin{equation}\label{eqn:rate}
\begin{split}
\frac{1}{T_2^*} \approx & \frac{1}{T_2^*\{\text{NV-}^{13}\mathrm C \}} + \frac{1}{T_2^*\{\text{NV-N}\}} + \frac{1}{T_2^*\{\text{other spins}\}} + \\ & \frac{1}{T_2^*\{\text{strain grad.}\}} +
\frac{1}{T_2^*\{\text{B-field grad.}\}} + \\ & \frac{1}{T_2^*\{\text{temp. fluctuations}\}} + ...,
\end{split}
\end{equation}
where $T_2^*\{\cdot \}$ describes the $T_2^*$-limit imposed by a particular dephasing mechanism, and the ``$\approx$''-symbol indicates that individual dephasing rates add approximately linearly.
}

DQ magnetometry employs the $\{-1,+1\}$ sub-basis of the NV spin$-1$ system for quantum sensing. In this basis, noise sources that shift the $|\pm 1\rangle$ states in common mode (e.g., strain inhomogeneities { and spectrum drifts due to temperature fluctuations of the host diamond;  fourth and sixth term in Eqn.\,\ref{eqn:rate}, respectively}) are suppressed by probing the energy difference between the $|+1\rangle$ and $|-1\rangle$ spin states. In addition, the NV DQ spin coherence accumulates phase due to an external magnetic field at twice the rate 
of traditional single quantum (SQ) coherence magnetometry, for which the $|0\rangle$ and $|+1\rangle$ (or $|-1\rangle$) spin states are probed. 
DQ magnetometry provides enhanced susceptibility to target magnetic field signals while also making the spin coherence twice as sensitive to magnetic noise, including interactions with the paramagnetic spin bath. 
We therefore use resonant radiofrequency control to decouple the bath spins from the NV sensors { (second and third term in Eqn.\,\ref{eqn:rate})}. By employing both DQ magnetometry and spin bath driving { with isotopically enriched samples}, we elucidate and effectively eliminate the dominant sources of NV spin ensemble dephasing, realizing up to a $16\times$ extension of the ensemble $T_2^*$ in diamond. These techniques are also compatible with Ramsey-based DC sensing, and we find up to an $8\times$ improvement in DC magnetic field sensitivity. Our results should enable broadband DC sensing using NV spin ensembles with spin interrogation times approaching those used in AC sensing; and may aid in the fabrication of optimized samples for a wide range of solid-state sensor species.


\subsection*{Double Quantum Magnetometry}

The enhanced sensitivity to magnetic fields and insensitivity to common-mode noise sources in this DQ basis can be understood by considering the full ground-state Hamiltonian for NV centers, given by (neglecting the hyperfine interaction)~\cite{Doherty2013},
\begin{equation}\label{eqn:NVHamiltonian}
\begin{split}
H/h = & D\,\mathbf S_z^2 + \frac{\gamma_{NV}}{2\pi} \mathbf B \cdot \mathbf S + M_z \mathbf S_z^2 + \\
& M_x \mathbf (\mathbf S_y^2 - \mathbf S_x^2) + M_y (\mathbf S_x \mathbf S_y + \mathbf S_y \mathbf S_x),
\end{split}
\end{equation}
where $D\approx 2.87\,$GHz is the zero-field spin-state splitting, $\mathbf S = \{\mathbf S_x, \mathbf S_y, \mathbf S_z\}$ are the dimensionless spin-1 operators, $\mathbf B = \{B_x, B_y, B_z\}$ are the local magnetic field components, $\gamma_{NV}/2\pi \approx 28\,$GHz/T is the NV gyromagnetic ratio, and $\{M_x, M_y, M_z\}$ describe the strain and electric field contributions to $H$~\cite{Barson2017}. Ignoring terms $\propto \mathbf S_x$, $\mathbf S_y$ due to the large zero-field splitting $D$ and a small applied bias $B_z \gtrsim 10\,$mT along $z$, the transition frequencies $f_{\pm1}$ (see Fig.\,\ref{fig:fig1}b) are
\begin{equation}\label{eqn:nvenergies}
	f_{\pm1} \approx D + M_z \pm \frac{\gamma_{NV}}{2\pi} B_z.
\end{equation}
On-axis strain contributions ($\propto\,M_z$) as well as temperature fluctuations  ($\frac{\partial D}{\partial T} \approx -74$\,kHz/K)~\cite{Acosta2010, Toyli2013}
%
%
shift the $f_{\pm1}$ transitions linearly.  Thus, when performing DQ magnetometry where the difference $\Delta f = f_{+1} -f_{-1}$ is probed, their effects are to first order suppressed. In addition, a pertubative analysis of the complete Hamiltonian in Eqn.\,\ref{eqn:NVHamiltonian} (see Suppl.\,VII) 
shows that the effects of off-axis strain contributions ($\propto M_x, M_y$) on DQ magnetometry are reduced by a factor $\sqrt{M_x^2 + M_y^2}/(\gamma_{NV} B_z/\pi)$, proportional to the bias magnetic field $B_z$. Similarly, the effects of off-axis magnetic fields ($\propto B_x, B_y$) on DQ magnetometry are suppressed due to the large zero-field splitting $D$, and are also largely common-mode.
Working in the DQ basis at moderate bias fields can therefore lead to an enhancement in $T_2^*$ for NV ensembles if strain inhomogeneities, small off-axis magnetic field gradients ($B_x, B_y \ll D$), or temperature fluctuations are significant mechanisms of inhomogeneous spin dephasing. This result should be contrasted with single NV measurements in which $T_2^*$ and $T_2$ in the DQ basis were found to be approximately half the values in the SQ basis, i.e., $\tau^\text{coh}_\text{DQ} \approx \tau^\text{coh}_\text{SQ}/2$~\cite{Fang2013,Mamin2014}. Since spatial inhomogeneities are not relevant for single centers, the reduced decay times are attributed to an increased sensitivity to magnetic noise in the DQ basis due to the paramagnetic spin bath.

For example, using vector magnetic microscopy (VMM)~\cite{Glenn2017}, we mapped the on-axis strain component $M_z$ in a $1\,$mm$^2$-region for one of the three NV ensemble diamond samples studied in this work ($[\text N]=0.75\,$ppm, Sample B) to quantify the length-scale and magnitude of strain inhomogeneity (Fig. \ref{fig:fig1}c). From this analysis, we estimate an average strain gradient $M_z/L \approx 2.8\,$kHz/$\upmu$m, which, as we show below, is in good agreement with the observed SQ $T_2^*$ in our samples.

\subsection*{Spin Bath Driving}

%
To mitigate NV spin dephasing due to the spin bath, we drive the bath electronic spins~\cite{DeLange2012, Knowles2014} using resonant radiofrequency (RF) radiation.
In Fig.\,\ref{fig:fig1}d, we display the spin resonance spectrum of a nitrogen-rich diamond sample ($[\text N]=0.75\,$ppm, Sample B), recorded via the NV double electron-electron resonance (DEER) technique~\cite{Slichter1990} in the frequency range 100 - 500\,MHz (see Suppl.\,IX). 
The data reveal 6 distinct spectral peaks attributed to $^{14}$N substitutional defects in the diamond lattice. The resonance peaks have an approximate amplitude ratio of 1:3:1:3:3:1 resulting from the four crystallographic Jahn-Teller orientations of the nitrogen defects at two possible angles with respect to an applied bias magnetic field ($B_z= 8.5\,$mT, aligned along the [111]-axis), as well as 3 hyperfine states~\cite{Ammerlaan1981,Davies1979,Davies1981} (see Suppl.\,IX for details). 
{Additional smaller peaks $i$ and $ii$ are attributed to dipole-forbidden nitrogen spin transitions and other electronic dark spins}~\cite{Yamamoto2013}.

In pulsed spin bath driving~\cite{DeLange2012}, a multi-frequency RF $\pi$-pulse is applied to each of the bath spin resonances midway through the NV Ramsey sequence, decoupling the bath from the NV sensor spins in analogy to a refocusing $\pi$-pulse in a spin echo sequence~\cite{DeLange2010}. Alternatively, the bath spins can be driven with continuous wave (CW)~\cite{DeLange2012, Knowles2014}. In this case, the Rabi drive strength $\Omega_\text{Bath}$ at each bath spin resonance frequency must significantly exceed the characteristic coupling strength $\gamma$ between the bath spins and NV centers, i.e., $\Omega_{Bath}/\gamma \gg 1$, to achieve effective decoupling. Under this condition, the baths spins undergo many Rabi oscillations during the characteristic dipolar interaction time $1/\gamma$. {As a result, the dipolar interaction with the bath is incoherently averaged and the NV spin dephasing time increases.}



\section*{Results}

We studied three diamond samples with increasing nitrogen concentrations that are summarized in Table\,\ref{tab:tab1}. Samples A ($[\text N] \lesssim 0.05\,$ppm) and B ($[\text N] = 0.75\,$ppm) each consist of a $^{14}$N-doped, $\approx 100\,\upmu$m-thick chemical-vapor-deposition (CVD) layer (99.99$\%~^{12}$C) deposited on top of a diamond substrate. Sample C ($[\text N] = 10\, $ppm) possesses a $ 40\,\upmu$m-thick, $^{15}$N-doped CVD layer (99.95$\%~^{12}$C) on a diamond substrate.
For all three samples, the nitrogen-limited NV dephasing times can be estimated from the average dipolar interaction strength between electronic spins giving $T_{2,\text{NV-N}}^* \approx 350\,\upmu$s, $23\,\upmu$s, and $2\,\upmu$s for Samples A, B, and C, respectively. Analysis and measurements suggest that the $^{13}$C nuclear spin bath limit to $T_2^*$ is $\approx 100\,\upmu$s for Samples A and B, and $\approx 20\,\upmu$s for Sample C (for details, see Suppl.\,V). 
All samples are unirradiated and the N-to-NV conversion efficiency 
is $\lesssim 1\%$. Contributions from NV-NV dipolar interactions to $T_2^*$ can therefore be neglected. { The parameter regime covered by Samples A, B, and C was chosen to best illustrate the efficacy of DQ coherence magnetometry and spin bath driving.

}
%
%
%
\begin{widetext}
{\onecolumngrid

\begin{table}[t]
  \centering
    \begin{tabular}{ccccccccccc}
    \hline
    \hline \\[-1.5ex]
    Sample & [\text N] & $^{13}$C & [NV] & $T_2^{\text{meas}}$ & $T_{2,\text{SQ}}^{*,\text{meas}}$ & $T_{2,\text{DQ}}^{*,\text{meas}}$ & $T_{2,\text{NV-N}}^{*,\text{est}}$ & $T_{2,\mathrm{NV-^{13}C}}^{*,\text{est}}$ & $T_{2,\mathrm{NV-(^{13}C+N)}}^{*,\text{est}}$ & $dM_z^{\text{meas}}/dL$ \\
           & (ppm) & (\%)  & (cm$^{-3})$  & ($\upmu$s)  & ($\upmu$s)  & ($\upmu$s)  & ($\upmu$s)  & ($\upmu$s)  & ($\upmu$s) & (MHz/$\upmu$m)\\[0.5ex]
\hline
\\[-1.5ex]
    A     &$\lesssim 0.05$ & 0.01  & $\sim 3 \times 10^{12}$ & $\gtrsim 630$   & $5-12$  & 34(2)    & 350   & 100   & 78   & n/a \\
    B     & 0.75  & 0.01  & $\sim 10^{14}$ & $250-300$ & $1-10$  & 6.9(5)    & 23    & 100   & 19  & 0.0028 \\
    C     & 10    & 0.05  & $\sim 6 \times 10^{15}$ & $15-18$    & $0.3-1.2$ & $0.60(2)$   & 2     & 20    & 2   & n/a \\
    \hline
    \hline
    \end{tabular}%
   \caption{
  {Characteristics of Samples A, B, and C.} The estimated values $T_2^{*,\text{est}}$ are calculated using the contributions of $^{13}$C and nitrogen spins as described in the main text. Reasonable agreement is found between the estimated $T_{2,\mathrm{NV-(^{13}C+N})}^{*,\text{est}}$ and twice the measured $T_{2,\text{DQ}}^{*,\text{meas}}$, consistent with the twice faster dephasing in the DQ basis. Values listed with a $\sim$ symbol are order-of-magnitude estimates. For all samples, $[\text{NV}]\ll[\text N]$ and NV contributions to $T_2^*$ can be neglected (1\,ppm $= 1.76 \times 10^{17}\,$cm$^{-3}$). 
 }
\label{tab:tab1}%
\end{table}%
\twocolumngrid}

\end{widetext}
We measured $T_2^*$ values in the SQ and DQ bases, denoted $T_{2,\text{SQ}}^*$ and $T_{2,\text{DQ}}^*$ from here on, by performing a single- or two-tone $\pi/2 - \tau - \pi/2$ Ramsey sequence, respectively (see inset Fig.\,\ref{fig:fig2}). In both instances, the observed Ramsey signal exhibits a characteristic stretched exponential decay envelope that is modulated by the frequency detunings of the applied NV drive(s) from the NV hyperfine transitions. We fit the data to the expression $C_0 \exp\left[-(\tau / T_2^*)^p\right] \sum_{i} \cos(2 \pi f_i (\tau-\tau_{0,i}))$, where the free parameters in the fit are the maximal contrast $C_0$ at $\tau=0$, dephasing time $T_2^*$, stretched exponential parameter $p$, time-offsets $\tau_{0,i}$, and (up to) three frequencies $f_{i}$ from the NV hyperfine splittings. The $p$ value provides a phenomenological description of the decay envelope, which depends on the specific noise sources in the spin bath as well as the distribution of individual resonance lines within the NV ensemble. For a purely magnetic-noise-limited spin bath, the NV ensemble decay envelope exhibits simple exponential decay ($p=1$)~\cite{Abragam1983,Dobrovitski2008}; whereas a non-integer p-value ($p \neq 1$) suggests magnetic and/or strain gradient-limited NV spin ensemble dephasing.%

\subsection*{Strain-dominated dephasing (Sample A: low nitrogen density regime)}

Experiments on Sample A ($[\text N]\lesssim 0.05\, $ppm, $^{14}$N) probed the low nitrogen density regime. In different regions of this diamond, the measured SQ Ramsey dephasing time varies between $T_{2,\text{SQ}}^* \simeq 5 - 12\,\upmu$s, with $1<p<2$. Strikingly, even the longest measured $T_{2,\text{SQ}}^*$ is $\sim 30\times$ shorter than the calculated $T_{2, \text{NV-N}}^*$ given by the nitrogen concentration of the sample ($\gtrsim 350\,\upmu$s, see Table\,\ref{tab:tab1}) and is approximately $10\times$ smaller than the expected SQ limit due to $0.01\%$ $^{13}$C spins ($\simeq 100\,\upmu$s). This discrepancy indicates that dipolar broadening due to paramagnetic spins is not the dominant NV dephasing mechanism. Indeed, the spatial variation in $T_{2,\text{SQ}}^*$ and low concentration of nitrogen and $^{13}$C spins suggests that crystal lattice strain inhomogeneity is the main source of NV spin ensemble dephasing in this sample. For the measured NV ensemble volume ($\sim 10^4\,\upmu\mathrm{m}^3$) and the reference strain gradient (Fig.\,\ref{fig:fig1}c), we estimate a strain gradient limited dephasing time of $\sim 6\,\upmu$s, in reasonable agreement with the observed $T_{2,\text{SQ}}^*$.
%
%
\begin{figure}[ht]
  \centering
  \includegraphics{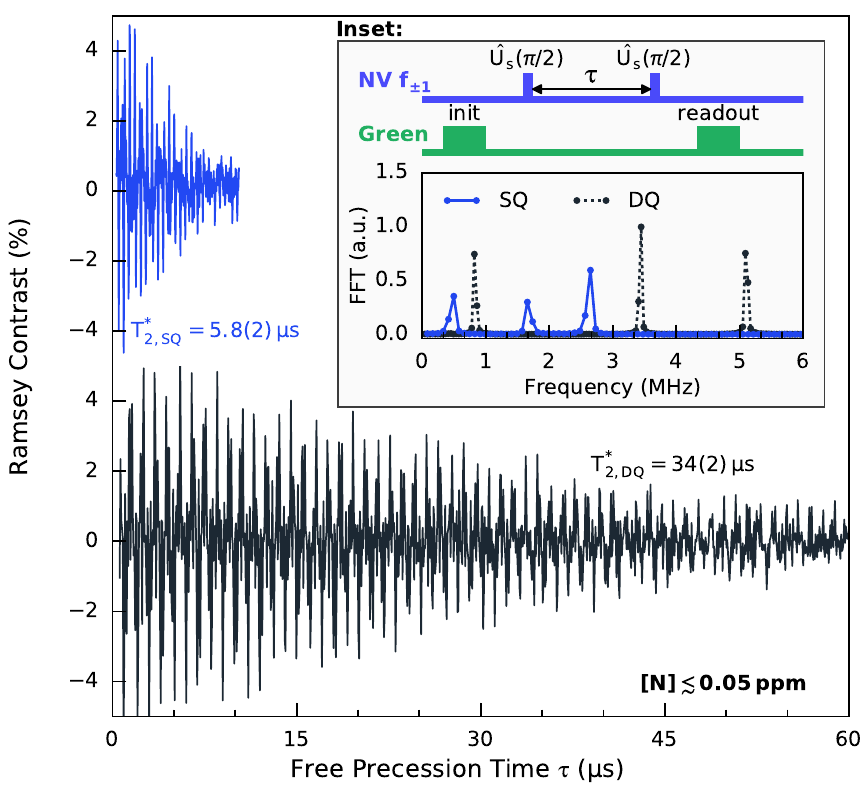}
  \caption{
  {NV Ramsey measurements for low nitrogen density sample (Sample A, ${[\text N] \lesssim 0.05\,}$ ppm) at an applied bias magnetic field of ${B_0 = 2.2\,}$mT.} Comparison of time-domain data and resulting fit values for the NV spin ensemble $T_2^*$ for the {single quantum (SQ) coherence, $\{0,+1\}$ (blue, upper); and the double quantum (DQ) coherence, $\{+1, -1\}$ (black, lower).} Upper inset: Illustration of DQ Ramsey protocol with two-tone microwave (MW) pulses, where $\hat{U}_{S=1}(\pi/2)$ is the spin-1 unitary evolution operator~\cite{Mamin2014}. For SQ measurements, a single-tone MW pulse is applied instead to generate the pseudo-spin-1/2 unitary evolution operator $\hat{U}_{S=1/2}(\pi/2)$. Lower inset: Discrete Fourier transform of the SQ (solid blue) and DQ (dashed black) Ramsey measurements with a MW drive detuned 0.4\,MHz from the $\{0,\pm 1\}$ transitions. NV sensor spins accumulate phase twice as quickly in the DQ basis as in the SQ basis.}
  \label{fig:fig2}
\end{figure}
Measurements in the DQ basis at moderate bias magnetic fields are to first order strain-insensitive, and therefore provide a means to eliminate the dominant contribution of strain to NV spin ensemble dephasing. Fig.\,\ref{fig:fig2} shows data for $T_{2}^*$ in both the SQ and DQ bases for an example region of Sample A with SQ dephasing time $T_{2,\text{SQ}}^* = 5.8(2) \,\upmu$s and $p = 1.7(2)$. For these measurements, we applied a small $2.2\,$mT bias field parallel to one NV axis (misalignment angle $<3\degree$) to lift the $|\pm 1\rangle$ degeneracy, and optimized the magnet geometry to reduce magnetic field gradients over the sensing volume (see Suppl.\,VI). 
In the DQ basis, we find $T_{2,\text{DQ}}^* = 34(2) \upmu$s with $p=1.0(1)$, which is a $ \sim 6\times$ improvement over the measured $T_{2}^*$ in the SQ basis. We observed similar $T_{2}^*$ improvements in the DQ basis in other regions of this diamond. Our results suggest that in the low nitrogen density regime, dipolar interactions with the $^{13}$C nuclear spin bath are the primary decoherence mechanism when DQ basis measurements are employed to remove strain and temperature effects.  Specifically, the measured $T_{2,\text{DQ}}^*$ and $p$ values in Sample A are consistent with the combined effect of NV dipolar interactions with (i) the $0.01\,\%$ concentration of $^{13}$C nuclear spins ($T_{2,\text{N-$^{13}$C}}^*/2 \simeq 50\,\upmu$s) and (ii) residual nitrogen spins $[\text N]\sim 0.05\,$ppm; with an estimated net effect of $T_{2,\text{DQ}}^* \simeq 39\,\upmu$s. Diamond samples with greater isotopic purity $(^{12}$C$ > 99.99\%)$ would likely yield further enhancements in $T_{2,\text{DQ}}^*$.

\subsection*{Strain- and dipolar-dominated dephasing (Sample B: intermediate nitrogen density regime)}

\begin{figure}[ht!]
  \centering
  \includegraphics{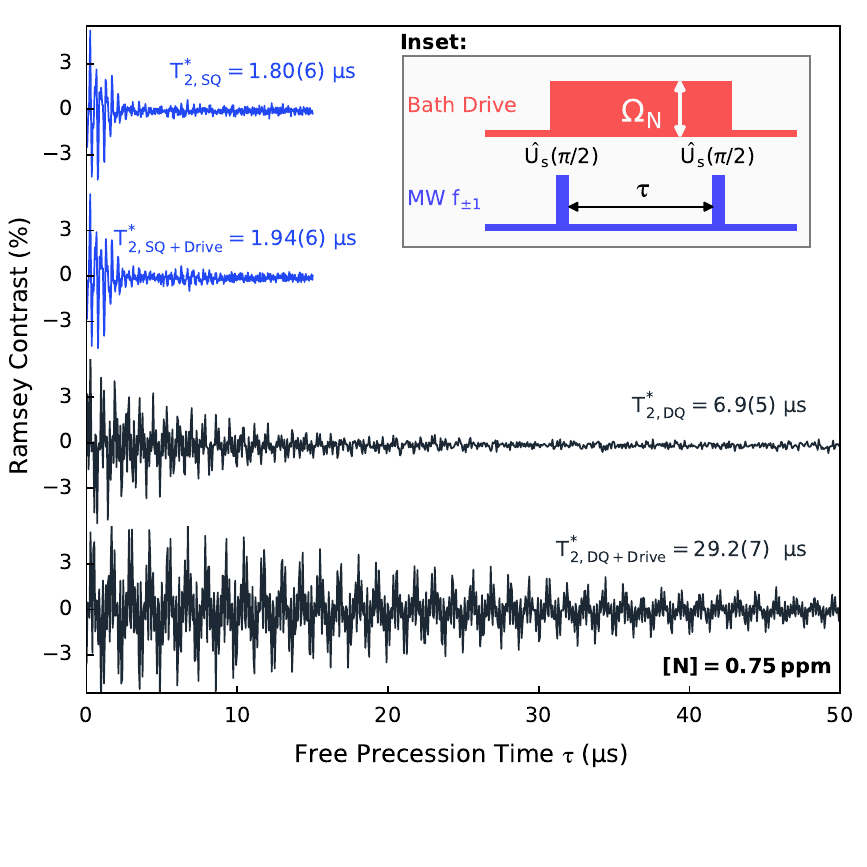}
  \caption{
  {NV Ramsey measurements for intermediate nitrogen density sample (Sample B, (${[\text N]=0.75\,}$ ppm) at an applied bias magnetic field of ${B_0 = 8.5\,}$mT.}
  Comparison of time-domain data and resulting fit values for the NV spin ensemble $T_2^*$ for the {single quantum (SQ) coherence $\{0,+1\}$ (blue, 1\textsuperscript{st} from top); the SQ coherence with spin-bath drive (blue, 2\textsuperscript{nd} from top); the DQ coherence with no drive (black, 3\textsuperscript{rd} from top); and the DQ coherence with spin-bath drive (black, 4\textsuperscript{th} from top). There is a $16.2\times$ improvement of $T_2^*$ with spin-bath drive when the DQ coherence is used for sensing compared to SQ with no drive. Inset: Two-tone NV Ramsey protocol with applied spin-bath bath drive resonant with nitrogen spins.} 
  }
  \label{fig:fig3}
\end{figure}

Although Sample B ($[\text N]=0.75\,$ppm, $^{14}$N) contains more than an order of magnitude higher nitrogen spin concentration than Sample A ($[\text N]\lesssim 0.05\, $ppm), we observed SQ Ramsey dephasing times $T_{2,\text{SQ}}^{*} \simeq 1 - 10\, \upmu$s in different regions of Sample B, which are similar to the results from Sample A. We conclude that strain inhomogeneities are also a significant contributor to NV spin ensemble dephasing in Sample B . Comparative measurements of $T_{2}^*$ in both the SQ and DQ bases yield a more moderate increase in $T_{2,\text{DQ}}^*$ for Sample B than for Sample A. Example Ramsey measurements of Sample B are displayed in Fig.\,\ref{fig:fig3}, showing $T_{2,\text{SQ}}^*$ = $1.80(6)\,\upmu$s in the SQ basis increasing to $T_{2,\text{DQ}}^*$ = $6.9(5)\,\upmu$s in the DQ basis, a $\sim 4\times$ extension. The observed $T_{2,\text{DQ}}^*$ in Sample B approaches the expected limit set by dipolar coupling of NV spins to residual nitrogen spins in the diamond ($T_{2,\text{N-NV}}^*/2 \simeq 12\,\upmu$s), but is still well below the expected DQ limit due to $0.01\,\%$ $^{13}$C nuclear spins ($\simeq 50\,\upmu$s).

%
%

Measuring NV Ramsey decay in both the SQ and DQ bases while driving the nitrogen spins, either via application of CW or pulsed RF fields~\cite{DeLange2012, Knowles2014}, 
is effective in revealing the electronic spin bath contribution to NV ensemble dephasing. With continuous drive fields of Rabi frequency $\Omega_N = 2\,$MHz applied to nitrogen spin resonances $1-6$, $i$, and $ii$ (see Fig. \ref{fig:fig1}d), we find that $T_{2,\text{SQ+Drive}}^* = 1.94(6)\,\upmu$s, which only marginally exceeds $T_{2,\text{SQ}}^* = 1.80(6)\,\upmu$s. This result is consistent with NV ensemble SQ dephasing being dominated by strain gradients in Sample B, rendering spin bath driving ineffective in the SQ basis. In contrast, DQ Ramsey measurements exhibit a significant additional increase in $T_2^*$ when the bath drive is applied, improving from $T_{2,\text{DQ}}^* = 6.9(5) \,\upmu$s to $T_{2,\text{DQ+Drive}}^* = 29.2(7)\,\upmu$s. This $\sim 16\times$ improvement over $T_{2,\text{SQ}}^*$ confirms that, for Sample B without spin bath drive, dipolar interactions with the nitrogen spin bath are the dominant mechanism of NV spin ensemble dephasing in the DQ basis. Note that the NV dephasing time for Sample B with DQ plus spin bath drive is only slightly below that for Sample A with DQ alone ($\approx 34\,\upmu$s). We attribute this $T_2^*$ limit in Sample B primarily to NV dipolar interactions with $0.01\%$ $^{13}$C nuclear spins. There is also an additional small contribution from magnetic field gradients over the detection volume ($\sim 10^4\,\upmu\mathrm{m}^3$) due to the four times larger applied bias field ($B_0$ = $8.5\,$mT), relative to Sample A, which was used in Sample B to resolve the nitrogen ESR spectral features (see Suppl.~Table\,S3 and S4). 
We obtained similar extensions of $T_2^*$ using pulsed driving of the nitrogen bath spins (see Supp.\,X).

We also characterized the efficacy of CW spin bath driving for increasing $T_2^*$ in both the SQ and DQ bases (see Fig.\,\ref{fig:fig4}a). While $T_{2,\text{SQ}}^*$ remains approximately constant with varying Rabi drive frequency $\Omega_{N}$, $T_{2,\text{DQ}}^*$ exhibits an initial rapid increase and saturates at $T_{2,\text{DQ}}^* \approx 27\,\upmu$s for $\Omega_{N} \gtrsim 1\,$MHz (only resonances $1-6$ are driven here). To explain the observed trend, we introduce a model that distinguishes between (i) NV spin ensemble dephasing due to nitrogen bath spins, which depends upon bath drive strength $\Omega_\text{N}$, and (ii) dephasing from drive-independent sources (including strain and $^{13}$C spins), 
\begin{equation}\label{eqn:t2stardrive}
1/T_{2}^* = 1/T_{2,\text{NV-N}}^*(\Omega_\text{N}) + 1/T_{2,\text{other}}^*.
\end{equation}
%
Taking the coherent dynamics of the bath drive into account (see Suppl.\,VIII), 
the data is well described by the functional form
\begin{equation} \label{eqn:t2stardrive2}
1/T_{2,\text{NV-N}}^*(\Omega_\text{N}) = \Delta m \times \gamma_\text{NV-N} \frac{\delta_\text{N}^2}{\delta_\text{N}^2 + \Omega_\text{N}^2},
\end{equation}
%
%
where $\Delta m = 1 (2)$ is the change in spin quantum number in the SQ (DQ) basis and $\delta_\text{N} = \gamma_\text{N}/2\pi$ is the Lorentzian linewidth (half width at half max) of the nitrogen spin resonances measured through DEER ESR (Fig.\,\ref{fig:fig1}d). Although we find that NV and nitrogen spins have comparable $T_2^*$ ($\gamma_\text{NV-N} \approx \gamma_\text{N}$, see Suppl.\,XI), 
the effective linewidth $\delta_\text{N}$ relevant for bath driving is increased due to imperfect overlap of the nitrogen spin resonances caused by a small misalignment angle of the applied bias magnetic field. 

Using the NV-N dipolar estimate for Sample B, $\gamma_\text{NV-N} \approx 2\pi \times 7\,$kHz, $\delta_\text{N} \approx 80\,$kHz extracted from DEER measurements (Suppl.\,XI), 
and a saturation value of $T_{2,\text{other}}^* \approx 27\,\upmu$s, we combine Eqns.\,\ref{eqn:t2stardrive} and\,\ref{eqn:t2stardrive2} and plot the calculated  $T_{2}^*$ as a function of $\Omega_\text{N}$ in Fig.\,\ref{fig:fig4}a (black, dashed line). The good agreement between the model and our data in the DQ basis suggests that Eqns.\,\ref{eqn:t2stardrive} and\,\ref{eqn:t2stardrive2} capture the dependence of $T_{2}^*$ on drive field magnitude (i.e., Rabi frequency). Alternatively, we fit the model to the DQ data (red, solid line) and extract $\gamma_\text{NV-N}^{fit} = 2\pi \times 9.3(2)$kHz and $\delta_\text{N}^{fit} = 60(3)\,$kHz, in reasonable agreement with our estimated parameters. In summary, the results from Sample B show that the combination of spin bath driving and sensing in the DQ basis suppresses inhomogeneous NV ensemble dephasing due to both interactions with the nitrogen spin bath and strain-gradients. Similar to Sample A, further enhancement in $T_{2}^*$ could be achieved with improved isotopic purity, as well as reduced magnetic-gradients due to the applied magnetic bias field.

%
%
\begin{widetext}
\onecolumngrid 

\begin{figure}[ht]
 \centering
  \includegraphics[scale=1]{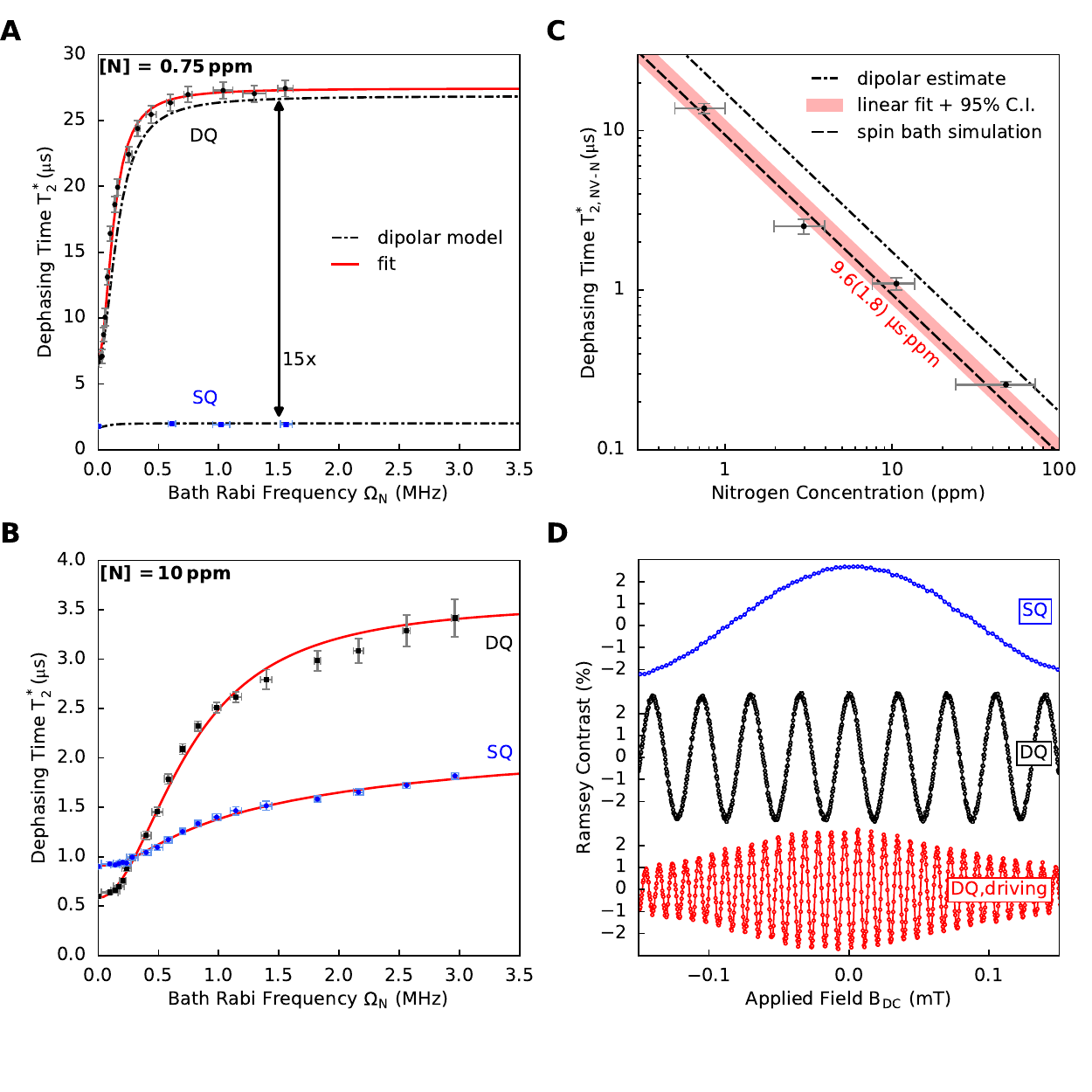}
  \caption{
   Application of quantum control techniques to extend NV spin ensemble dephasing time (${T_2^*}$) and increase DC magnetic field sensitivity. (a) Ramsey measurements of $T_2^*$ in the single quantum (SQ, blue) and double quantum (DQ, black) bases for different spin-bath drive strengths (Rabi frequencies) for Sample B ($\mathrm [\text N] = 0.75\,$ppm) at $B_0 = 8.5\,$mT. Black dashed-dotted line is calculated from a model of NV spins that are dipolar-coupled to a multi-component spin bath (Eqn. \ref{eqn:t2stardrive}). The red solid line is a fit of the model to the $T_2^*$ data (see main text for details). {(b)} Same as (a) but for Sample C ($[\text N]=10\,$ppm) and $B_0 = 10.3\,$mT. {(c)} Measured $T_{2,\text{N-NV}}^* \equiv 2\times T_{2,\text{DQ}}^*$ as a function of nitrogen concentration for Samples B, C, D, E. {Samples were selected to have a predominately electronic nitrogen (P1) spin bath using DEER ESR measurements.} The black dashed-dotted line is the dipolar-interaction-estimated dependence of $T_2^*$ on nitrogen concentration (Suppl.\,V). 
We fit the data using an orthogonal-distance-regression routine to account for the uncertainties in [N] and $T_2^*$. A fit to the form $1/T_2^* = A_\text{NV-N} [\text N]$ yields $A_\text{N-NV} = 2\pi \times 16.6(2.6)\,$kHz/ppm [$1/A_\text{NV-N} = 9.6(1.8)\,\upmu \mathrm{s}\,\cdot\,$ppm]. The red shaded region indicates the 95\,\% standard error of the fit value for $A_\text{N-NV}$. The black dashed line is the expected scaling extracted from numerical simulations using a second-moment analysis of the NV ensemble ESR linewidth (see text for details). {(d)} Measured Ramsey DC magnetometry signal $S \propto C \sin(\phi(\tau))$ for Sample B, in the SQ and DQ bases, as well as the DQ {sub-basis} with spin-bath drive (see main text for details). There is a $36\times$ faster oscillation in the DQ {sub-basis} with spin-bath drive compared to SQ with no drive. This greatly enhanced DC magnetic field sensitivity is a direct result of the extended $T_2^*$, with the sensitivity enhancement given by $2\times \sqrt{\tau_\text{DQ+Drive}/\tau_\text{SQ}}$ at equal contrast. The slight decrease in observed contrast in the DQ + drive case for $|B_{DC}| > 0.05\,$mT is a result of changes in the Zeeman resonance frequencies of the nitrogen spins due to the applied test field $B_{DC}$, which was not corrected for in these measurements.
  }
  \label{fig:fig4}
\end{figure}
\twocolumngrid

\end{widetext}

\subsection*{Dipolar-dominated dephasing (Sample C: high nitrogen density regime)}

Spin bath driving results for Sample C  ($[\text N]=10\,$ppm, $^{15}$N) are shown in Fig.\,\ref{fig:fig4}b. At this high nitrogen density, interactions with the nitrogen bath dominate NV spin ensemble dephasing, and $T_{2,\text{SQ}}^*$ and $T_{2,\text{DQ}}^*$ both exhibit a clear dependence on spin bath drive strength $\Omega_\text{N}$. With no drive ($\Omega_\text{N}=0$), we measured $T_{2,\text{DQ}}^* \approx T_{2,\text{SQ}}^*/2$, in agreement with dephasing dominated by a paramagnetic spin environment and the twice higher precession rate in the DQ basis~\cite{Fang2013, Mamin2014,MacQuarrie2015}. Note that this result is in contrast to the observed DQ basis enhancement of $T_{2}^*$ at lower nitrogen density for Samples A and B (Figs.\,\ref{fig:fig2} and\,\ref{fig:fig3}). We also find that $T_2^*$ in Sample C increases more rapidly as a function of spin bath drive amplitude in the DQ basis than in the SQ basis, such that $T_{2,\text{DQ}}^*$ surpasses $T_{2,\text{SQ}}^*$ with sufficient spin bath drive strength. We attribute the $T_2^*$-limit in the SQ basis ($\simeq 1.8\,\upmu$s) to strain inhomogeneities in this sample, whereas the longest observed $T_2^*$ in the DQ basis ($\simeq  3.4\,\upmu$s) is in agreement with dephasing due to the $0.05\%\,^{13}$C and 0.5\,ppm residual $^{14}$N spin impurities. The latter were incorporated during growth of this $^{15}$N sample (see Suppl.\,Table\,S5).

In Fig.\,\ref{fig:fig4}c we plot $T_{2,\text{NV-N}}^* \equiv 2\times T_{2,\text{DQ}}^*$ versus sample nitrogen concentration $[\text N]$ {to account for the twice faster dephasing of the DQ coherence.} To improve the range of $[\text N]$ coverage, we include DQ data for additional diamonds, Samples D ($[\text N] = 3\,$ppm) and E ($[\text N]= 48\,$ppm). To our knowledge, the dependence of the NV spin ensemble dephasing time on $[\text N]$ has not previously been experimentally reported. Fitting the data to the function $1/T_{2,\text{NV-N}}^* = A_\text{NV-N}\cdot[\text N]$ (red shaded region), we find the characteristic NV-N interaction strength for NV ensembles to be $A_\text{NV-N} = 2\pi \times 16.6(2.6)\,$kHz/ppm [$1/A_\text{N-NV} = 9.6(1.8)\,\upmu\mathrm{s}\cdot\mathrm{ppm}$] {in the SQ sub-basis}. This value is about $1.8\times$ larger than the dipolar-estimate $\gamma_{\text{e-e}} = 2\pi \times 9.1\,$kHz/ppm (black dashed-dotted line), which is used above in estimates of NV dephasing due to the nitrogen spin bath. We also performed numerical spin bath simulations for the NV-N spin system and determine the second moment of the dipolar-broadened single NV ESR linewidth~\cite[Ch. III and IV]{Abragam1983}. By simulating $10^4$ random spin bath configurations, we extract the ensemble-averaged dephasing time from the distribution of the single NV linewidths~\cite{Dobrovitski2008}. The results of this simulation (black dashed line) are in excellent agreement with the experiment and confirm the validity of our obtained scaling for $T_{2,\text{NV-N}}^*(\mathrm N)$. Additional details of the simulation are provided in Ref.~\cite{Bauch2017}.

\subsection*{Ramsey DC Magnetic Field Sensing}

We demonstrated that combining the two quantum control techniques can greatly improve the sensitivity of Ramsey DC magnetometry. Fig.\,\ref{fig:fig4}d compares the accumulated phase for SQ, DQ, and DQ plus spin bath drive measurements of a tunable static magnetic field of amplitude $B_{DC}$, for Sample B. Sweeping $B_{DC}$ leads to a characteristic observed oscillation of the Ramsey signal $S \propto C \sin(\phi)$, where $C = C_0 \exp[-\left(\tau/T_2^*\right)^p]$ is the measurement contrast and $\phi = \Delta m \times \gamma_{NV} B_{DC} \tau$ is the accumulated phase during the free precession interval $\tau \approx  T_2^*$. Choosing $\tau_{\text{SQ}}=1.308\,\upmu$s and $\tau_\text{DQ+Drive}=23.99\,\upmu$s (see Suppl.\,XII), 
we find a $36.3(1.9)\times$ faster oscillation period (at equal measurement contrast) when DQ and spin bath driving are both employed, compared to a SQ measurement. This enhancement in phase accumulation, and hence DC magnetic field sensitivity, agrees well with the expected improvement ($2 \times \tau_\text{DQ+Drive}/\tau_\text{{SQ}}$ = 36.7).

\section*{Discussion}

Our results (i) characterize the dominant spin dephasing mechanisms for NV ensembles in bulk diamond (strain and interactions with the paramagnetic spin bath); and (ii) demonstrate that the combination of DQ magnetometry and spin bath driving can greatly extend the NV spin ensemble $T_2^*$.  For example, in Sample B we find that these quantum control techniques, when combined, provide a $16.2\times$ improvement in $T_2^*$. Operation in the DQ basis protects against common-mode inhomogeneities and enables an extension of $T_2^*$ for samples with $[\text N] \lesssim 1$\,ppm. In such samples, strain inhomogeneities are found to be the main causes of NV spin ensemble dephasing. In samples with higher N concentration ($[\text N] \gtrsim 1$\,ppm), spin bath driving in combination with DQ sensing provides an increase of the NV ensemble $T_2^*$ by decoupling paramagnetic nitrogen and other electronic dark spins from the NV spins. 
Our results suggest that quantum control techniques may allow the NV ensemble $T_{2}^*$ to approach the bare Hahn echo coherence time $T_2$.  Note that spin bath driving may also be used to enhance the NV ensemble $T_2$ in Hahn echo, dynamical decoupling~\cite{DeLange2010,Pham2012}, and spectral decomposition experimental protocols~\cite{Bar-Gill2012}.

Furthermore, we showed that the combination of DQ magnetometry and spin bath driving allows improved DC Ramsey magnetic field sensing. The relative enhancement in photon-shot-noise-limited sensitivity (neglecting experimental overhead time) is quantified by $2 \times\sqrt{\zeta}$, where the factor of two accounts for the enhanced gyromagnetic ratio in the DQ basis and $\zeta \equiv T_{2,\text{DQ}}^*/T_{2,\text{SQ}}^*$ is the ratio of maximally achieved $T_2^*$ in the DQ basis (with spin bath drive when advantageous) and non-driven $T_2^*$ in the SQ basis. For Samples A, B, and C, we calculate $2 \times \sqrt{\zeta} = 5.2\times$, $8.1\times$, and $3.9\times$, respectively, using our experimental values. In practice, increasing $T_{2}^*$ also decreases the fractional overhead time associated with NV optical initialization and readout, resulting in even greater DC magnetic field sensitivity improvements and an approximately linear sensitivity enhancement with $\zeta$ (see Suppl.\,XII). 
We expect that these quantum control techniques will remain effective when integrated with other approaches to optimize NV ensemble magnetic field sensitivity, such as high laser power and good N-to-NV conversion efficiency. In particular, conversion efficiencies of $1-30\,\%$ have been reported for NV ensemble measurements~\cite{Acosta2010,Wolf2015,Grezes2015,Barry2016}, such that the nitrogen spin bath continues to be a relevant spin dephasing mechanism.

There are multiple avenues for further improvement in NV ensemble $T_2^*$ and DC magnetic field sensitivity, beyond the gains demonstrated in this work. First, the $^{13}$C limitation to $T_2^*$, observed for all samples, can be mitigated via improved isotopic purity ([$^{12}$C] $>99.99\,\%$); or possibly through driving of the nuclear spin bath \cite{London2013}. Second, more efficient RF delivery will enable faster spin bath driving (higher Rabi drive frequency $\Omega_\text{N}$), which will be critical for decoupling denser nitrogen baths and thereby extending $T_2^* \propto \Omega_\text{N}^2/\delta_{\text N}^2 \propto \Omega_\text{N}^2/[\text N]^2$ (see Eqn. \ref{eqn:t2stardrive2}). Third, short NV ensemble $T_2^*$ times have so far prevented effective utilization of more exotic readout techniques, e.g., involving quantum logic~\cite{Jiang2009,Neumann2010,Lovchinsky2016} or spin-to-charge-conversion~\cite{Shields2015,Jaskula2017}. Such methods offer greatly improved NV spin-state readout fidelity but introduce substantial overhead time, typically requiring tens to hundreds of microseconds per readout operation. The NV spin ensemble dephasing times demonstrated in this work ($T_2^*\gtrsim 20\,\upmu$s) may allow effective application of these readout schemes, which only offer sensitivity improvements when the sequence sensing time (set by $T_2^*$ for DC sensing) is comparable to the added overhead time.
We note that the NV ensemble $T_2^*$ values obtained in this work are the longest for any electronic solid-state spin system at room temperature (see comparison Fig.\,S2) suggesting that state-of-the-art DC magnetic field sensitivity~\cite{Barry2016,Chatzidrosos2017} may be increased to $\sim 100\,$fT/$\sqrt{\text{Hz}}$ for optimized NV ensembles in a diamond sensing volume $\sim (100\, \upmu$m)$^3$ (see discussion on NV ensemble DC magnetic field sensitivity optimization in Barry et al.~\cite{Barry2016}). In conclusion, DQ magnetometry in combination with spin bath driving allows for order-of-magnitude increase in the NV ensemble $T_2^*$ in diamond, providing a clear path to ultra-high sensitivity DC magnetometry with NV ensemble coherence times approaching $T_2$.


\bibliography{references}

\section*{Acknowledgements}

We thank David Le Sage for his initial contributions to this project. We thank Joonhee Choi, Soonwon Choi, and Renate Landig for fruitful discussions. This material is based upon work supported by, or in part by, the United States Army Research Laboratory and the United States Army Research Office under Grant No. W911NF1510548; the National Science Foundation Electronics, Photonics and Magnetic Devices (EPMD), Physics of Living Systems (PoLS), and Integrated NSF Support Promoting Interdisciplinary Research and Education (INSPIRE) programs under Grants No. ECCS-1408075, PHY-1504610,  and EAR-1647504, respectively; and Lockheed Martin under award A32198. This work was performed in part at the Center for Nanoscale Systems (CNS), a member of the National Nanotechnology Coordinated Infrastructure Network (NNCI), which is supported by the National Science Foundation under NSF award no. 1541959. CNS is part of Harvard University. P. K. acknowledges support from the Intelligence Community Postdoctoral Research Fellowship Program. J. M. S. was supported by a Fannie and John Hertz Foundation Graduate Fellowship and a National Science Foundation (NSF) Graduate Research Fellowship under Grant 1122374.


\section*{Author contributions statement}

E. B., C. A. H., J. M. S., M. J. T., J. F. B., and R. L. W. conceived the experiments,  C. A. H. and E. B. conducted the experiments and analyzed the results.  P. K. provided the strain analysis. E. B. and S. S. provided the spin bath simulation. All authors contributed to and reviewed the manuscript.  R. L. W. supervised the work.

\include{suppl}

\end{document}

%% file: suppl.tex








\clearpage
\onecolumngrid

\begin{center}

\textbf{\large Supplemental Materials: Ultralong dephasing times in solid-state spin ensembles via quantum control}
\end{center}

\setcounter{equation}{0}
\setcounter{figure}{0}
\setcounter{table}{0}
\setcounter{page}{1}


\renewcommand{\theequation}{S\arabic{equation}}
\renewcommand{\thefigure}{S\arabic{figure}}
\renewcommand{\thetable}{S\arabic{table}}

\title{Supplemental Materials: Ultralong dephasing times in solid-state spin ensembles via quantum control}

\author{Erik Bauch}
\affiliation{Department	of	Physics,	Harvard	University,	Cambridge,	Massachusetts	02138,	USA}

\author{Connor A. Hart}
\affiliation{Department	of	Physics,	Harvard	University,	Cambridge,	Massachusetts	02138,	USA}

\author{Jennifer M. Schloss}
\affiliation{Department	of	Physics,	Massachusetts	Institute	of	Technology,	Cambridge,	Massachusetts	02139,	USA}
\affiliation{Department	of	Physics,	Harvard	University,	Cambridge,	Massachusetts	02138,	USA}
\affiliation{Center	for	Brain	Science,	Harvard	University,	Cambridge,	Massachusetts	02138, USA}

\author{Matthew J. Turner}
\affiliation{Department	of	Physics,	Harvard	University,	Cambridge,	Massachusetts	02138,	USA}
\affiliation{Center	for	Brain	Science,	Harvard	University,	Cambridge,	Massachusetts	02138, USA}

\author{John F. Barry}
\affiliation{Department	of	Physics,	Harvard	University,	Cambridge,	Massachusetts	02138,	USA}
\affiliation{Lincoln Laboratory, Massachusetts Institute of Technology, Lexington, Massachusetts 02420, USA}

\author{Pauli Kehayias}
\affiliation{Department	of	Physics,	Harvard	University,	Cambridge,	Massachusetts	02138,	USA}
\affiliation{
Harvard-Smithsonian Center	for	Astrophysics,	Cambridge,	Massachusetts	02138,	USA}

\author{Swati Singh}
\affiliation{Williams College,
Department of Physics,
33 Lab Campus Drive,
Williamstown, Massachusetts 01267}

\author{Ronald L. Walsworth}
\email{rwalsworth@cfa.harvard.edu}
\affiliation{Department	of	Physics,	Harvard	University,	Cambridge,	Massachusetts	02138,	USA}
\affiliation{Center	for	Brain	Science,	Harvard	University,	Cambridge,	Massachusetts	02138, USA}
\affiliation{
Harvard-Smithsonian Center	for	Astrophysics,	Cambridge,	Massachusetts	02138,	USA}

\maketitle

\tableofcontents

\section{Experimental methods}\label{sec:expmethods}

A custom-built, wide-field microscope collected the spin-dependent fluorescence from an NV ensemble onto an avalanche photodiode. Optical initialization and readout of the NV ensemble was accomplished via 532\,nm continuous-wave (CW) laser light focused through the same objective used for fluorescence collection (Fig.\,1a). 
The detection volume was given by the 532\,nm beam excitation at the surface (diameter $\approx 20\, \upmu$m) and sample thickness (100$\, \upmu$m for Samples A and B, 40$\, \upmu$m for Sample C). A static magnetic bias field was applied to split the $|-1\rangle$ and $|+1\rangle$ degeneracy in the NV ground state using two permanent samarium cobalt ring magnets in a Helmholtz-type configuration, with the generated field aligned along one [111] crystallographic axis of the diamond ($\equiv \hat{z}$). The magnet geometry was optimized using the Radia software package \cite{Elleaume1998} to minimize field gradients over the detection volume (see Suppl.\,\ref{sec:gradient}). A planar waveguide fabricated onto a glass substrate delivered $2-3.5\,$GHz microwave radiation for coherent control of the NV ensemble spin states. To manipulate the nitrogen spin resonances (see Fig.\,1d), 
a 1$\,$mm-diameter copper loop was positioned above the diamond sample to apply $100 - 600\,$MHz radiofrequency (RF) signals, synthesized from up to eight individual signal generators. Pulsed measurements on the NV and nitrogen spins were performed using a computer-controlled pulse generator and microwave switches.\\
%
%
\begin{figure}[ht]
  \centering
  \includegraphics[width=0.75\textwidth]{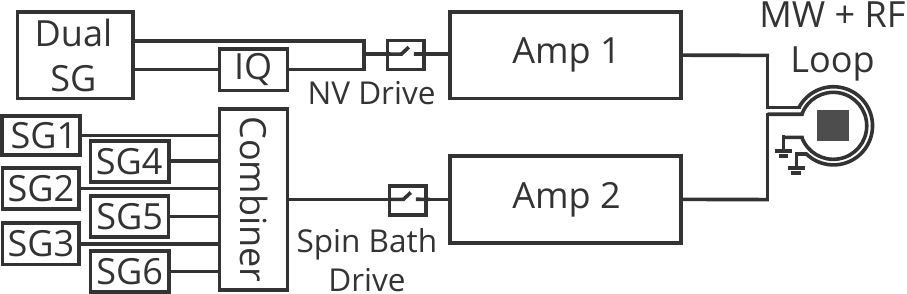}  
  \caption{{Microwave generation and delivery schematic}.
  For NV spin state control: Single and two-tone signals are generated using a dual channel Windfreak Technology Synth HD signal generator. One channel includes a Marki IQ-1545 mixer to manipulate the relative phase between both channels.  A single Minicircuits ZASWA-2-50DR+ switch is used to generate the NV control pulses before amplification with a Minicircuits ZHL-16W-43 amplifer. The NV control fields are delivered to the diamond sample  using a fabricated microwave waveguide (diameter $500 \, \upmu$m). For spin bath control: Up to eight single channel Windfreak Technology Synth NV signal generators are combined before passing through a switch and a Minicircuits ZHL-100W-52 100 W amplifier. The amplified field is delivered via a grounded cooper loop ($1\,$mm diameter).}
  \label{fig:supplsetup}
\end{figure}
%

The NV ESR measurement contrast (Fig.\,2,\,3,\,and\,4d) 
is determined by comparing the fluorescence from the NV ensemble in the $|0\rangle$ state (maximal fluorescence) relative to the $|+1\rangle$ or $|-1\rangle$ state (minimal fluorescence)~\cite{Doherty2013} and is defined as visibility $C = \mathrm{\frac{max - min}{max+min}}$. The DEER (Fig.\,1d) 
and DC magnetometry contrast (Fig.\,4d) 
are calculated in the same fashion, but are reduced by $\approx 1/e$ since the best phase sensitivity in those measurements is obtained at $\tau \approx T_2$  and $\tau \approx T_2^*$, respectively (see Suppl.\,\ref{sec:nvp1linewidth} and \ref{sec:dcsense}). For noise rejection, most pulse sequences in this work use a back-to-back double measurement scheme~\cite{Bar-Gill2013}, where the accumulated NV spin ensemble phase signal is first projected onto the $|0\rangle$ state and then onto the $|+ 1\rangle$ (or $|- 1\rangle$) state. The contrast for a single measurement is then defined as the visibility of both sequences.

\clearpage 

\section{Sample information (all samples)}

Information for all samples used in this study is summarized in Table\,\ref{tab:sampleinfoall}.

\begin{table}[htbp!]
  \centering
   \caption{{Detailed information for Samples A - E.} Values with $\sim$ symbol are order-of-magnitude estimates. For all samples, $[\text{NV}]\ll[\text N]$ and NV contributions to $T_2^*$ can be neglected (1\,ppm $= 1.76 \times 10^{17}\,$cm$^{-3}$).}.
    \begin{tabular}{ccccccccccc}
    \hline
    \hline
    Sample & [\text N] & $^{13}$C & [NV] & $T_2^{\text{meas}}$ & $T_{2,\text{SQ}}^{*,\text{meas}}$ & $T_{2,\text{DQ}}^{*,\text{meas}}$ & $T_{2,\text{NV-N}}^{*,\text{est}}$ & $T_{2,\text{NV-13C}}^{*,\text{est}}$ & $T_{2,\text{NV-(13C+N)}}^{*,\text{est}}$ & $dM_z^{\text{meas}}/dL$ \\
           & (ppm) & (\%)  & (cm$^{-3})$  & ($\upmu$s)  & ($\upmu$s)  & ($\upmu$s)  & ($\upmu$s)  & ($\upmu$s)  & ($\upmu$s) & (MHz/$\upmu$m)\\
\hline \\[-1.5ex]
    A     &$\lesssim 0.05$ & 0.01  & $\sim 3 \times 10^{12}$ & $\gtrsim 630$   & $5-12$  & 34    & 350   & 100   & 78   & n/a \\
    B     & 0.75  & 0.01  & $\sim 10^{14}$ & $250-300$ & $1-10$  & 14    & 23    & 100   & 19  & 0.0028 \\
    C     & 10    & 0.05  & $\sim 6 \times 10^{15}$ & 15-18    & $0.3-1.2$ & 0.6   & 2     & 20    & 2   & n/a \\
    \hline \\[-1.5ex]
    D     & 3     & $< 0.01$ & $\sim 5 \times 10^{15}$ & 53   & 0.4   & 1.3   & 6     & 100   & $>6$     & n/a \\
    E     & 48    & 1.1   & $\sim 1 \times 10^{17}$ & 1.8   & 0.07  & 0.12  & 0.3   & 1     & 0.2   & n/a \\
\hline
\hline
    \end{tabular}%
  \label{tab:sampleinfoall}%
\end{table}%

\section{Survey of dephasing times}

\begin{figure}[htbp!]
  \centering
  \includegraphics[width=0.65\textwidth]{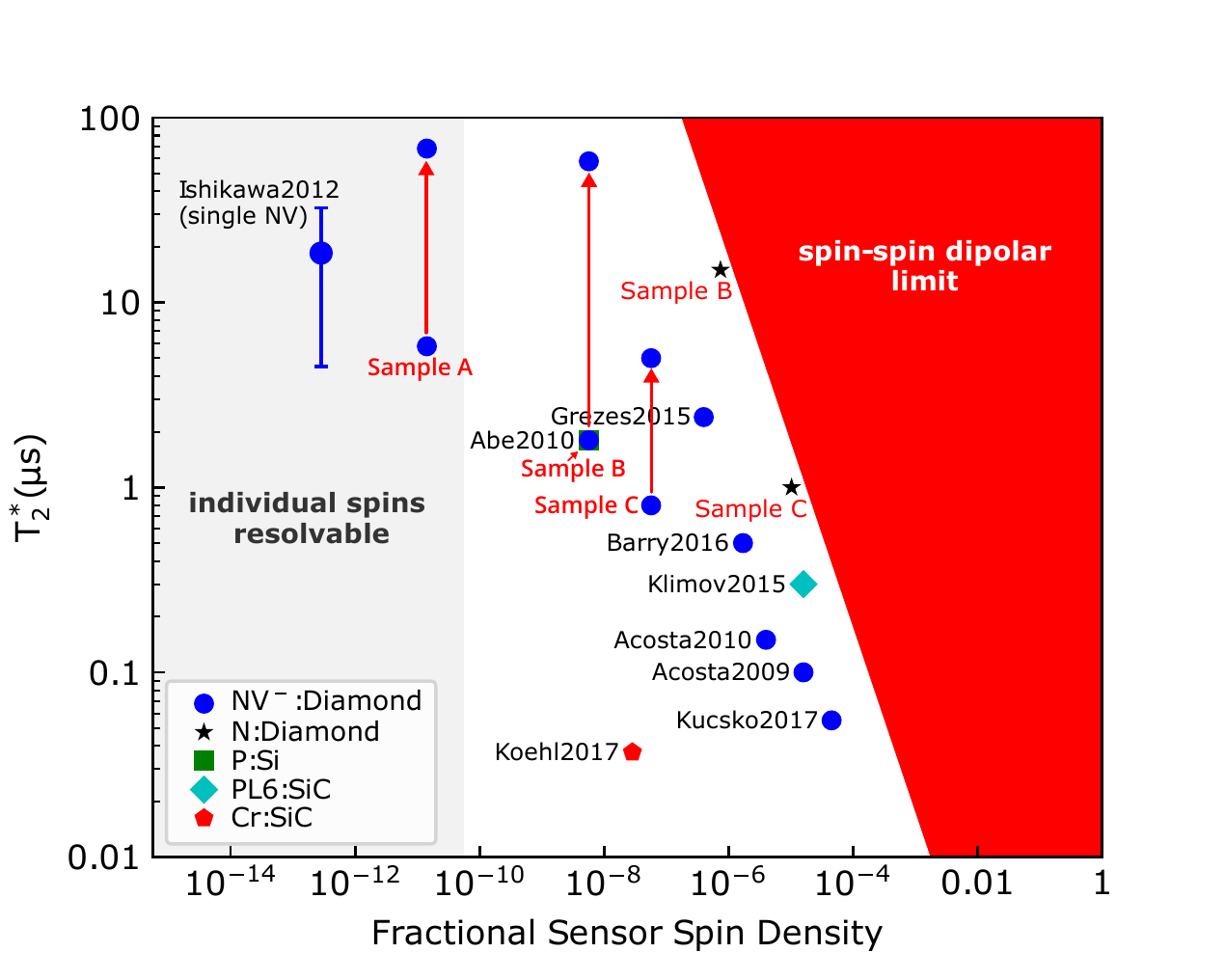}  
  \caption{
  Inhomogeneous spin dephasing times. Experimental results from this work are compared to that of related spin defect systems (see legend). Inhomogeneous dephasing due to paramagnetic bath spins (e.g., nitrogen and $^{13}$C nuclear spins in diamond), strain fields and other effects limit ${T_2^*}_\text{ens}$ at lower sensor-spin densities $\ll 1$. At higher sensor-spin densities approaching unity, spin-spin interaction places an upper bound on the ensemble dephasing time (red shaded area). This limit to $T_2^*$ is estimated using $\gamma_{e-e}$ (see Suppl.\,\ref{sec:suppl_spinbathcontrib}) and a fractional sensor-spin density of 1 corresponds to $\sim 10^{23}$\,cm$^{-3}$. Red arrows indicate improvement from the bare $T_2^*$ as measured in the NV SQ basis and increase when DQ sensing and spin bath drive (where advantageous) are employed to suppress inhomogeneities. The maximal obtained $T_2^*$ values for Sample A, B and C are multiplied by a factor of two to account for the twice higher gyromagnetic ratio in the NV DQ basis. 
%
The region in which individual, single spins are resolvable with confocal microscopy ($\sim 200\,$nm average spin separation) is shown in gray. $T_2^*$ values determined in Sec. \ref{sec:nvp1linewidth} for the nitrogen spins (P1 centers) are shown for reference as well (black stars). Measurements by Ishikawa et al. were performed on single NV centers and error bars indicate the spread in measured $T_2^*$ values. {\ The following references were used: Abe2010~\cite{Abe2010}, Acosta2009~\cite{Acosta2009}, Acosta2010~\cite{Acosta2010}, Barry2016~\cite{Barry2016}, Grezes2015~\cite{Grezes2015}, Ishikawa2012~\cite{Ishikawa2012}, Klimov2015~\cite{Klimov2015}, Koehl2017~\cite{Koehl2017}, and Kucsko2017~\cite{Kucsko2017}. }
  }
  \label{fig:surveyt2star}
\end{figure}

In Fig.\,\ref{fig:surveyt2star} we show a survey of inhomogeneous dephasing times for electronic solid-state spin ensembles.

\section{Strain contribution to $T_2^*$}\label{sec:straincontrib}

The on-axis strain component $M_z$ in Sample B was mapped across a $1\times 1\,$mm area using a separate wide-field imager of NV spin-state-dependent fluorescence. A bias field $B_0 \sim1.5\,$mT was applied to split the spin resonances from the four NV orientations. 
Measurements were performed following the vector magnetic microscopy (VMM) technique~\cite{Glenn2017}. Eqn.\,3
in the main text was used to analyze the measured NV resonance frequencies from each camera pixel (ignoring $M_x$ and $M_y$ terms as small perturbations, see Suppl. \ref{sec:nvham}).  This procedure yielded the average $B_x$, $B_y$, and $B_z$ magnetic field components, as well as the $M_{z}$ on-axis strain components for all four NV orientations in each camera pixel, corresponding to 2.42\,$\upmu$m$\times2.42\,\upmu$m transverse resolution on the diamond sample. 
Figure\,1c 
of the main text shows the resulting map of the on-axis strain inhomogeneity $M_z$ in Sample B for the NV orientation interrogated in this work. This map indicates an approximate strain gradient of 2.8 kHz/$\upmu$m across the field of view. The estimated strain gradient was used for all samples, while recognizing the likely variation between samples and within different regions of a sample.  Across a 20-$\upmu$m diameter spot, the measured strain inhomogeneity corresponds to a $T_{2}^*$ limit of $\approx 6\,\upmu$s, which compares well with the measured variation in $T_{2,\text{SQ}}^*$ for Samples A and B (see Table 1). Note that the contributions to $M_z$ can be microscopic (e.g., due {\ to nearby point defects}) or macroscopic (e.g., due to crystal defects with size $>10 \, \upmu$m). In addition, the VMM technique integrates over macroscopic gradients within the {\ depth of field of the VMM microscope. For the present experiments the resolution along the z-axis (i.e., perpendicular to the diamond surface) is given approximately by the thickness of the NV-diamond layer}. Consequently, the strain gradient estimate shown in Fig.\,1c 
is a measure of $M_z$ gradients in-plane within the NV layer, and strain gradients across the NV layer thickness are not resolvable in this measurement. 

\section{Spin bath contribution to $T_2^*$}\label{sec:suppl_spinbathcontrib}

The NV spin ensemble $T_2^*$ as a function of nitrogen concentration is estimated from the average dipolar coupling between electronic nitrogen spins, which is given by $\gamma_{e-e} = a \times \dfrac{\mu_0}{4\pi} g^2\mu_B^2/ \hbar \dfrac{1}{ \langle r\rangle^3} \approx 2\pi \times 9.1 \cdot [\text N] \,$kHz/ppm, where $\mu_0$ is the vacuum permeability, $g$ is the electron g-factor, $\mu_B$ is the Bohr magneton, $\hbar$ is the reduced Planck constant, $\langle r\rangle = 0.55 [\text N]^{-1/3}$ is the average spacing between electronic nitrogen spins as a function of density $[\text N]$ (in parts-per-million) within diamond~\cite{Hoch1988}, and $a$ is a factor of order unity collecting additional parameters from the dipolar estimate such as the angular dependence and spin resonance lineshape of the ensemble~\cite{Abragam1983}. A sample with $[\text N]= 1$\,ppm has an estimated $T_{2,\text{NV-N}}^* \approx 1/(2\pi\times 9.1\,\mathrm{kHz}) = 17.5\,\upmu$s using this dipolar estimate. Similarly, Table \ref{tab:sampleinfoall} gives the estimates $T_{2,\text{NV-N}}^*$ for Samples A, B, and C.\\

Uncertainties in nitrogen concentration $[\text N]$ used in Fig.\,4c 
are estimated by considering: the values reported by the manufacturer (Element Six Inc.); fluorescence measurements in a confocal microscope (Sample A); and Hahn echo $T_2$ measurements using the calibration value $T_2 (\text N) \simeq 165\,\upmu$s\,$\cdot\,$ppm reported in Ref.~\cite{Bauch2017} (Samples B and C). For example, for Sample B, Element Six reports $[\text N] = 1\,$ppm, whereas the measured $T_2=300\,\upmu$s suggests $[\text N] = 0.5\,$ppm. The average value is thus used: $[\text N] = 0.75 \pm 0.25$\,ppm.\\

\begin{figure}[ht]
  \centering
  \includegraphics[width=0.5\textwidth]{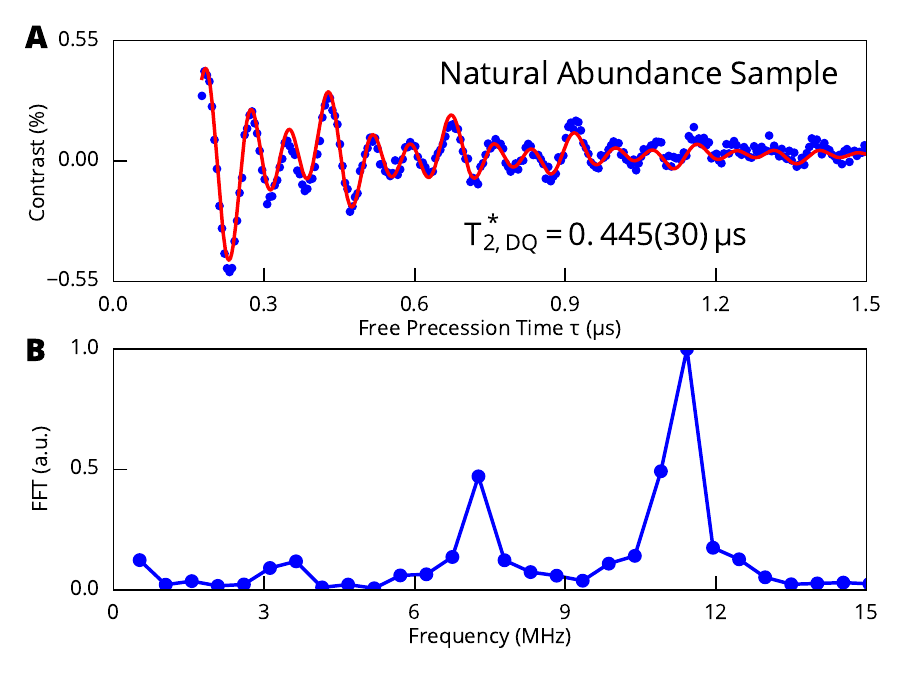}
  \caption{{NV Ramsey measurement for natural isotope abundance diamond sample.} {(a)} DQ Ramsey measurement on a natural abundance sample ($[\text N] \simeq 0.4\,$ppm, $[^{13}\mathrm C] = 1.07\,\%$) yields $T_{2,\text{DQ}}^*\,=\, 0.445(30) \, \upmu$s. {(b)} Fourier transform of Ramsey signal showing the enhanced precession in the DQ basis. A frequency detuning from the center hyperfine state of $3.65\,$MHz was chosen in this measurement; by sensing in the DQ basis, the detuning from each hyperfine state has acquired a factor of two.}
  \label{fig:13Cramsey}
\end{figure}

In the dilute $^{13}$C limit ($n_{{13}\text C} \lesssim 1.1\,\%$, where $n_{\text{${13}$C}}$ is the $^{13}$C spin concentration in percent), the NV-$^{13}$C contact interaction can be neglected and thus the NV ensemble ESR linewidth is expected to be linearly-dependent on the $^{13}$C concentration~\cite{Abragam1983,Abe2010}, i.e., $1/T_{2,\text{NV-13C}}^* = A_{\text{NV-}^{13}\text C} \cdot n_{\text{${13}$C}}$. An NV spin ensemble $T_2^*$ measurement on a natural abundance sample with $n_{\text{${13}$C}} = 1.07\,\%$ therefore provides a reasonable lower-bound estimate for $A_{\text{NV-${13}$C}}$ from which the $^{13}$C contribution in our diamond samples can be calculated. Fig.\,\ref{fig:13Cramsey} shows a DQ Ramsey measurement of a natural $^{13}$C abundance sample. Via a fit to the Ramsey data in the time domain, we extract $T_{2,\text{DQ}}^* = 445(30)$\,ns and $p = 1.0(1)$. After correcting for the small contribution of 0.4\,ppm nitrogen spins in the sample using the calibration found in Fig.\,4c 
of the main text, we calculate $A_{\text{NV-13C}} \approx 2\pi \times 160\,$kHz/\% (1/$A_{\text{$^{13}$C}} \approx 1\,\upmu\mathrm{s}\,\cdot\%$) from which we determine the NV-$^{13}$C limits given in Table 1 and the main text of the paper. 

\section{Magnetic field gradient contribution to $T_2^*$}
\label{sec:gradient}
\begin{figure}[ht!]
  \centering
  \includegraphics[width=0.75\textwidth]{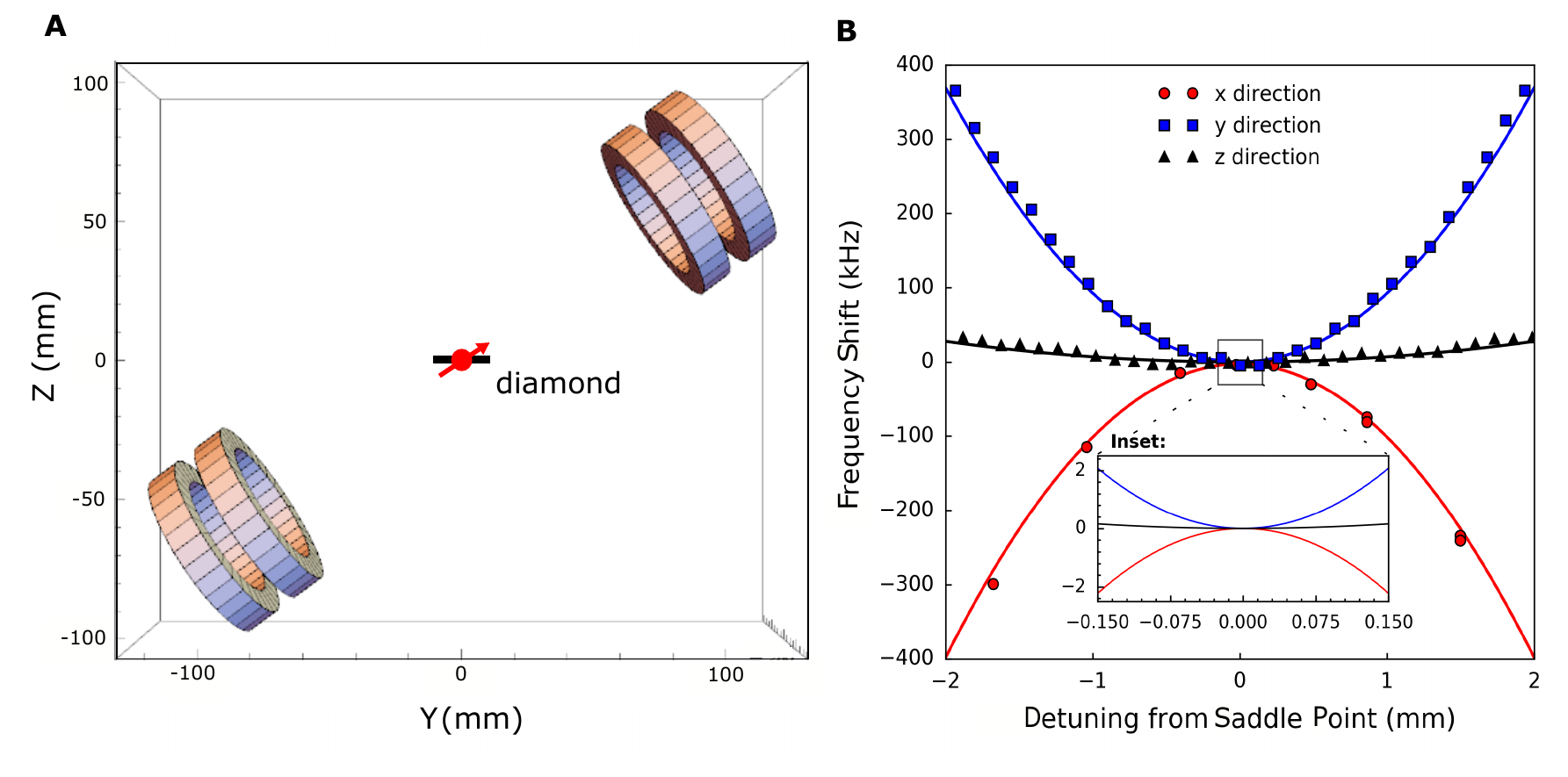}
  \caption{{Design of homogeneous magnetic bias field.}
  {(a)} Magnet geometry used to apply an external $B_0$ field along one NV orientation within the diamond crystal (typically [111]) as modeled using Radia~\cite{Elleaume1998}. Red arrow depicts the NV orientation class interrogated in these experiments; black rectangle represents diamond sample approximately to scale. {(b)} Magnets are translated along three axes to measure the $B_0$ field strength (shift in ESR transition frequency) as a function of detuning from the origin $(x,y,z =0)$ where the origin is defined as the center of the collection volume. Solid lines depict Radia simulation results while plotted points correspond to measured values. Inset: Zoomed-in view for length scale relevant for NV fluorescence collection volumes used in this work.}
\label{fig:bgradient}
\end{figure}

The NV-diamond epifluorescence microscope employs a custom-built samarium-cobalt (SmCo) magnet geometry designed to apply a homogeneous external field $B_0$ parallel to NVs oriented along the [111] diamond crystallographic axis. The field strength can be varied from 2 to 20\,mT (Fig.\,\ref{fig:bgradient}a). SmCo was chosen for its low reversible temperature coefficient (-0.03$\,\%/K$). Calculations performed using the Radia software package~\cite{Elleaume1998} enabled the optimization of the geometry to minimize $B_0$ gradients across the NV fluorescence collection volume. This collection volume is approximately cylindrical, with a measured diameter of $\approx 20\,\upmu$m and a length determined by the NV layer thickness along the z-axis ($40-100\,\upmu$m, depending on the diamond sample; see descriptions in the main text).\\

To calculate the expected $B_0$ field strength along the target NV orientation, the dimensions and properties of the magnets were used as Radia input, as well as an estimated $3\degree$ misalignment angle of the magnetic field with the NV axis. We find good agreement between the calculated field strength and values extracted from NV ESR measurements in Sample B, over a few millimeter lengthscale. The simulation results and measured values are plotted together in Fig.\,\ref{fig:bgradient}b. The z-direction gradient is reduced compared to the gradient in the xy-plane due to a high degree of symmetry along the z-axis for the magnet geometry.\\ 

Using data and simulation, we calculate that the $B_0$ gradient at $8.5\,$mT induces an NV ensemble ESR linewidth broadening of less than $0.1\,$kHz across the collection volume of Sample B. This corresponds to a $T_2^*$-limit on the order of $1\,$ms. However, due to interaction of the bias magnetic field with nearby materials and the displacement of the collection volume from the magnetic field saddle point, the experimentally realized gradient for Sample B was found to contribute an NV ESR linewidth broadening $\approx 1\,$kHz (implying a $T_2^*$-limit $\approx 320\,\upmu$s), which constitutes a small but non-negligible contribution to the $T_2^*$ values measured in this work. Ramsey measurements for Sample A were taken at a four times smaller bias field; we estimate therefore $\approx4\times$ better magnetic field homogeneity. 
For Sample C, with a layer thickness of $40 \, \upmu$m, the contribution of the magnetic field gradient at $10\,$mT to $T_2^*$ was similar to that of Sample B.


\section{NV Hamiltonian in single and double quantum bases} \label{sec:nvham}

In this section we discuss the influence of strain and magnetic fields in the single quantum (SQ) and double quantum (DQ) bases by considering several limiting cases. We first discuss how common-mode noise sources, i.e., sources that shift the NV $|-1\rangle$ and $|+1\rangle$ energy levels in-phase and with equal magnitude, are suppressed in the DQ basis. We then discuss how off-NV-axis strain fields are suppressed even by moderate bias magnetic fields. Lastly, we discuss the effect of off-axis magnetic fields on the NV spin-state energy levels and $T_2^*$. We begin with the negatively-charged NV ground electronic state electronic spin ($S=1$) Hamiltonian, which is given by~\cite{Doherty2013} {\   (neglecting hyperfine and quadrupolar effects):}
\begin{equation}\label{eqn:NVHam1}
\mathbf H/h = D \mathbf S_z^2 + \frac{\gamma_\text{NV}}{2\pi} (B_x \mathbf S_x + B_y \mathbf S_y + B_z \mathbf S_z) + M_z \mathbf S_z^2 + M_x \mathbf (\mathbf S_y^2 - \mathbf S_x^2) + M_y (\mathbf S_x \mathbf S_y + \mathbf S_y \mathbf S_x),
\end{equation}
where $D \approx 2.87\,$GHz is the NV zero-field splitting due to spin-spin interactions, $\mathbf \{B_x, B_y, B_z\}$ are the magnetic field components, $\{M_x, M_y, M_z\}$ collect strain and electric field components, $\{\mathbf S_x, \mathbf S_y, \mathbf S_z\}$ are the dimensionless spin-1 operators, and $\frac{g \mu_B}{h} = \frac{\gamma_\text{NV}}{2\pi} \approx 28.025\,$GHz/T is the NV gyromagnetic ratio.
Using $M_\perp \equiv -(M_x + i M_y)$, $B_\perp \equiv \frac{1}{\sqrt{2}}(B_x + i B_y)$, and the standard definitions for the spin operators $\{\mathbf S_x, \mathbf  S_y, \mathbf  S_z\}$, Eqn.\,\ref{eqn:NVHam1} reads in matrix form:
\begin{equation}\label{eqn:NVHam2}
\mathbf H/h = \left( \begin{array}{ccc}
D + M_z +\frac{\gamma_\text{NV}}{2\pi} B_z & \frac{\gamma_\text{NV}}{2\pi} B_\perp^* & M_\perp \\
\frac{\gamma_\text{NV}}{2\pi} B_\perp & 0 & \frac{\gamma_\text{NV}}{2\pi} B_\perp^* \\
M_\perp^* & \frac{\gamma_\text{NV}}{2\pi} B_\perp & D + M_z - \frac{\gamma_\text{NV}}{2\pi} B_z
\end{array} \right).
\end{equation}

\subsection*{Case 1: Zero strain, zero off-axis magnetic field}

For zero strain/electric field ($\{M_x, M_y, M_z\} = 0$) and zero off-axis magnetic field ($B_\perp = 0$), the Hamiltonian in Eqn.\,\ref{eqn:NVHam2} is diagonal:
\begin{equation}\label{eqn:NVHam3}
\mathbf H_0/h = \left( \begin{array}{ccc}
D + \frac{\gamma_\text{NV}}{2\pi} B_z &0 & 0 \\
0 & 0 & 0 \\
0 & 0 & D - \frac{\gamma_\text{NV}}{2\pi} B_z
\end{array} \right),
\end{equation}
and the energy levels are given by the zero-field splitting $D$ and Zeeman energies $\pm \frac{\gamma_\text{NV}}{2\pi} B_z$, 
\begin{equation} \label{eqn:NVener}
	E_{|\pm 1, 0\rangle}/h = \{D \pm \frac{\gamma_\text{NV}}{2\pi} B_z,0\},
\end{equation}
where $|\pm 1, 0\rangle$ are the Zeeman eigenstates
\begin{equation}
|+1\rangle  = \left( \begin{array}{c}
1 \\
0 \\
0
\end{array} \right), |-1\rangle  =
\left( \begin{array}{c}
0 \\
0 \\
1
\end{array} \right),\text{and } 
|0\rangle  = \left( \begin{array}{c}
0 \\
1 \\
0
\end{array} \right).
\end{equation}
NV spin ensemble measurements in the DQ basis, for which the difference between the $f_{-1} = E_{|0\rangle \rightarrow |-1>}$ and $f_{+1} = E_{|0\rangle \rightarrow |+1>}$ transitions is probed (see Fig.\,1b),  
are to first-order insensitive to inhomogeneities and fluctuations in $D$ (e.g., due to drift in temperature), and other common-mode noise sources. However, DQ measurements are twice as sensitive to magnetic fields along $B_z$. The DQ basis therefore provides both enhanced magnetic field sensitivity and protection against common-mode noise sources (for higher order effects see, e.g., the Supplement of Ref.~\cite{Fang2013}).

\subsection*{Case 2: Non-zero strain, zero off-axis magnetic field}

For non-zero strain/electric field components, but negligible off-axis magnetic fields ($B_\perp \approx 0$), the energy eigenvalues of the NV Hamiltonian (Eqn. \ref{eqn:NVHam2}) for the $|\pm 1\rangle$ states become
\begin{align}
E_{|\pm1\rangle}/h &= D + M_z \pm \sqrt{(\frac{\gamma_\text{NV}}{2\pi} B_z)^2 +  \lvert\lvert M_{\perp} \rvert\rvert^2}\\ \label{eqn:NVener2}
&\approx  D + M_z \pm 
\left[\frac{\gamma_\text{NV}}{2\pi} B_z 
+ \frac{\lvert\lvert M_{\perp} \rvert\rvert^2}{2 \frac{\gamma_\text{NV}}{2\pi} B_z}
+ \mathcal{O}\left(\frac{(\lvert\lvert M_{\perp} \rvert\rvert^4}{B_z^2} \right)
\right].
\end{align}
From Eqn.\,\ref{eqn:NVener2} it follows that off-axis strain ($\propto \lvert\lvert M_{\perp} \rvert\rvert$) is suppressed by moderate on-axis bias fields by a factor $\frac{\vert \vert M_\perp \vert \vert}{\gamma_\text{NV} B_z/\pi}$, as noted in the main text. Reported values for $\lvert\lvert M_{\perp} \rvert\rvert$ are $\sim 10\,$kHz~\cite{Fang2013} and $\sim 100\,$kHz~\cite{Jamonneau2015} for single NV centers in bulk diamond, and $\sim 7\,$MHz in nano-diamonds~\cite{Jamonneau2015}. Fig.\,\,1c 
in the main text shows that the measured on-axis strain $M_z$ in Sample B varies by $2-3\,$MHz (see Suppl.\,\ref{sec:straincontrib} for details).
\subsection*{Case 3: Non-zero off-axis magnetic field}
For non-zero off-axis magnetic field $(B_\perp \neq 0)$ we find the energy values for the NV Hamiltonian (Eqn.\,\ref{eqn:NVHam1}) by treating $B_\perp$ as a small perturbation, with perturbation Hamiltonian $\mathbf V \equiv \mathbf H - \mathbf  H_0$. To simplify the analysis we set $M_{||} = M_\perp = 0$. Using time-independent perturbation theory (TIPT, see for example Ref.~\cite{Sakurai2014}), the corrected energy levels are then given by
$E_{|\pm1,0 \rangle} \approx E_{|\pm1,0 \rangle}^{(0)} + E_{|\pm1,0 \rangle}^{(1)} + E_{|\pm1,0 \rangle}^{(2)} + \dots$
, where $E^{(0)}_{|\pm1,0 \rangle}$ are the bare Zeeman energies as given in Eqn. \ref{eqn:NVener} and $E_{|\pm1,0 \rangle}^{(k)}$ for $k>0$ are the k-th order corrections. The energy corrections at first and second order are:
\begin{align}
 E_{|\pm1,0 \rangle}^{(1)} & = \langle \pm 1, 0 | \mathbf  V | \pm1, 0 \rangle = 0,
 \end{align}
\centerline{and}
 \begin{align}
  E_{|\pm1\rangle}^{(2)} & = 
 \frac{\| \langle \mp 1| \mathbf  V | \pm 1 \rangle \|^2}{E_{|\pm1\rangle} 
 - E_{|\mp 1\rangle}} + 
 \frac{\|\langle 0| \mathbf  V | \pm 1 \rangle\|^2}{E_{|\pm1\rangle}} = \frac{\| \frac{\gamma_\text{NV}}{2\pi} B_\perp \|^2}{D \pm \frac{\gamma_\text{NV}}{2\pi} B_z} \approx \frac{\| \frac{\gamma_\text{NV}}{2\pi} B_\perp \|^2}{D},\\
 E_{|0\rangle}^{(2)} & = 
 \frac{\| \langle +1| \mathbf  V | 0 \rangle \|^2}{
 - E_{|+1\rangle}} + 
 \frac{\|\langle -1| \mathbf  V | 0 \rangle\|^2}{- E_{|-1\rangle}} = -\left(\frac{\|\frac{\gamma_\text{NV}}{2\pi} B_\perp \|^2}{D + \frac{\gamma_\text{NV}}{2\pi} B_z} + \frac{\| \frac{\gamma_\text{NV}}{2\pi} B_\perp \|^2}{D -\frac{\gamma_\text{NV}}{2\pi} B_z} \right) \approx - \frac{2 \| \frac{\gamma_\text{NV}}{2\pi} B_\perp \|^2}{D},
 \end{align}

where we have used in the last two lines the fact that $\frac{\gamma_\text{NV}}{2\pi} B_z \ll D$ in our experiments. The new transition frequencies for $E_{|0\rangle \rightarrow |\pm1 \rangle}$ are then found to be
\begin{equation} \label{eqn:NVener3}
f_{\pm1} \approx D + \frac{3 \| \frac{\gamma_\text{NV}}{2\pi} B_\perp \|^2}{D} \pm \frac{\gamma_\text{NV}}{2\pi} B_z.
\end{equation}
From Eqn.\,\ref{eqn:NVener3} it follows that energy level shifts due to perpendicular magnetic fields are mitigated by the large zero-field splitting $D$; and are further suppressed in the DQ basis, as they add (approximately) in common-mode. At moderate bias fields, $B_z = 2 - 20\,$mT, and typical misalignment angles of $\theta \sim 3\degree$ (or lower), we estimate a frequency shift of $0.1 - 1\,$kHz in the SQ basis.

\section{Spin bath driving model}\label{sec:drivig_model}
The effective magnetic field produced by the ensemble of nitrogen spins is modeled as a Lorentzian line shape with spectral width $\delta_\text{N}$ (half width at half max) and a maximum $\gamma_\text{NV-N}$ at zero drive frequency ($\Omega_\text{N} = 0$). This lineshape is derived in the context of dilute dipolar-coupled spin ensembles using the methods of moments~\cite[Ch. III and IV]{Abragam1983} and is consistent with NV DEER linewidth measurements (see Suppl.\,\ref{sec:nvp1linewidth}). The limit to the NV ensemble $T_2^*$ taking the bath drive into account is given by (see Eqn.\,5 of main text)
\begin{equation} \label{eqn:t2stardrive2methods}
1/T_{2,\text{NV-N}}^*(\Omega_\text{N}) = \Delta m \times \gamma_\text{NV-N} \frac{\delta_\text{N}^2}{\delta_\text{N}^2 + \Omega_\text{N}^2}. 
\end{equation}
At sufficiently high drive strengths ($\Omega_\text{N} \gg \delta_\text{N}$), the nitrogen spin ensemble is coherently driven and the resulting magnetic field noise spectrum is detuned away from the zero-frequency component, to which NV Ramsey measurements are maximally sensitive\,\cite{Cywinski2008}. For this case, the NV spin ensemble $T_2^*$ increases $ \propto \Omega_\text{N}^2/\delta_\text{N}^2$. At drive strength $\Omega_\text{N} \lesssim \delta_\text{N}$, however, the nitrogen spin ensemble is inhomogeneously driven and the dynamics of the spin bath cannot be described by coherent driving. Nonetheless, $1/T_2^*$ given by Eqn.\,\ref{eqn:t2stardrive2methods}
approaches $\gamma_\text{NV-N}$ in the limit $\Omega_\text{N} \rightarrow 0$, which is captured by the Lorentzian model.\\

This model (Eqn.\,\ref{eqn:t2stardrive2methods}) 
is in excellent agreement with the data for Sample B ([\text N] = 0.75\,ppm, $\delta_\text{N} \approx 11\,$kHz), for which $\Omega_\text{N} > \delta_\text{N}$ for the range of drive strengths employed. $\Omega_\text{N} > \delta_\text{N}$ also holds when the slight mismatch of nitrogen spin resonances is taken into account, effectively increasing the nitrogen linewidth relevant for bath driving ($\delta_\text{N} \approx 60\,$kHz, see discussion in main text). For Sample C ([\text N] = 10\,ppm, $\delta_\text{N} \approx 150\,$kHz), we find that the effective linewidth $\delta_\text{N}$ extracted from fitting the data in Fig.\,4b 
is about $4\times$ larger ($\approx 600$\,kHz) than what is expected from the dipolar estimate even after account for the small $B_0$ misalignment angle and resultant slight mismatch of nitrogen spin resonance frequencies. We attribute this discrepancy to incoherent dynamics at drive strength $\Omega_\text{N} \sim \delta_\text{N}$. Indeed, we find that for Sample C at drive strengths $\Omega_\text{N} \lesssim \delta_\text{N}$ the Ramsey signals exhibit multi-exponential decay with slow and fast decay rates, consistent with a larger effective $\delta_\text{N}$. To nonetheless enable a qualitative comparison with Sample B, in these instances the stretched exponential parameter is restricted to $p\geq1$ when extracting the NV spin ensemble $T_2^*$. At drive frequencies $\Omega_\text{N} > \delta_\text{N}$, the observed Ramsey signal returns to a simple exponential decay, confirming the validity of our driving model in this regime for Sample C. A more complete driving model, beyond the scope of this work, should take into account the changes of spin bath dynamics at drive strengths $\Omega_\text{N} \sim \delta_\text{N}$.



\section{$^{14}$N and $^{15}$N double electron-electron resonance spectra}\label{sec:nspectra}

\begin{figure}[ht]
  \centering
  \includegraphics[width=0.65\textwidth]{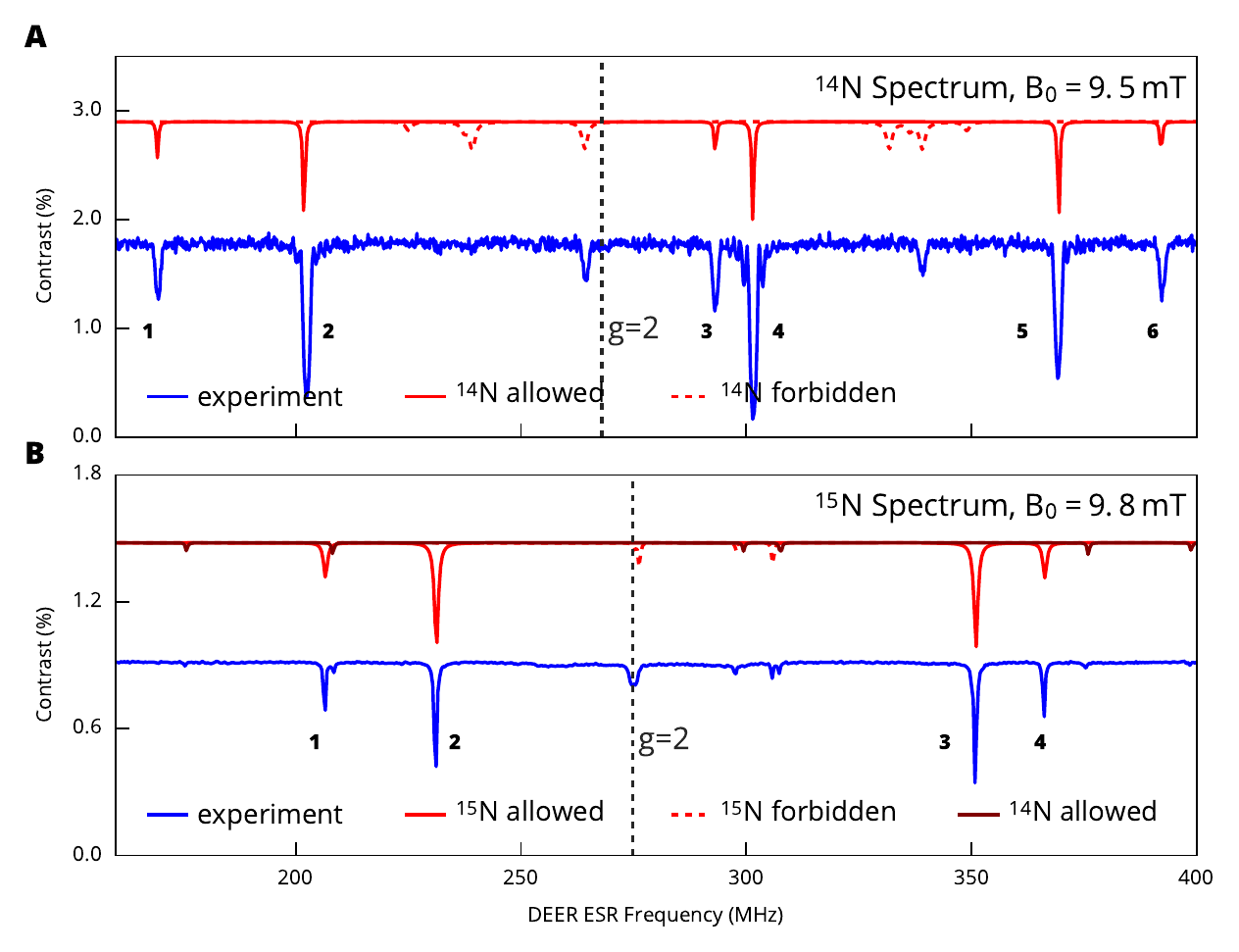}
  \caption{{$^{14}$N and $^{15}$N DEER spectra}. {(a)} Simulated (red) and measured (blue) $^{14}$N DEER spectra for Sample B ($[\text N] = 0.75\,$ppm, $^{14}$N). Dipole-allowed nitrogen hyperfine transitions are labeled 1 - 6. Smaller peaks are attributed to degenerate forbidden hyperfine transitions ($\Delta m_I \neq 0$) of the off-axis nitrogen orientations. Frequencies are simulated using Eqn.\,\ref{eqn:p1hamil2} and plotted as Lorentzians with widths and amplitudes chosen to reflect the experimental data. Allowed hyperfine transitions have an approximate amplitude ratio of 1:3:1:3:3:1 (see main text). The Larmor frequency of an electronic spin without hyperfine shift ($g=2$) is indicated as dashed black line. {(b)} Simulated (red) and measured (blue) $^{15}$N DEER spectrum for Sample C ($[\text N] = 10\,$ppm, $^{15}$N). Dipole-allowed hyperfine transitions are labeled 1 - 4. Smaller peaks are attributed to forbidden $^{15}$N hyperfine transitions and $g=2$ dark spins. The spectrum of a small abundance of $^{14}$N spins ($\approx 5\%$ of $^{15}$N density) is visible as well. 
  }
  \label{fig:suppldeern14andn15}
\end{figure}

We account for the $^{14}$N and $^{15}$N spin resonances, observed in NV double electron-electron resonance (DEER) spectra (see Fig.\,1d\,and \ref{fig:suppldeern14andn15}), 
in terms of Jahn-Teller, hyperfine, and quadrupolar splittings. The relevant spin Hamiltonian for the substitutional nitrogen defect is given by~\cite{Smith1959,Cook1966,Loubser1978,Cox1994}
\begin{equation} \label{eqn:p1hamil}
H_\text{P1}/h= \mu_B/h ~\mathbf B \cdot \mathbf g \cdot \mathbf S + \mu_\text{N}/h~\mathbf B \cdot \mathbf I + \mathbf S \cdot \mathbf A \cdot \mathbf I + \mathbf I \cdot \mathbf Q \cdot \mathbf I
\end{equation}
where $\mu_B$ is the Bohr magneton, $h$ is the Planck's constant, $\mathbf B =(B_x, B_y, B_z)$ is the magnetic field vector, $\mathbf g$ is the electronic g-factor tensor, $\mu_\text{N}$ is the nuclear magneton, $\mathbf S = (\mathbf S_x, \mathbf S_y, \mathbf S_z)$ is the electronic spin vector, $\mathbf A$ is the hyperfine tensor, $\mathbf I = (\mathbf I_x, \mathbf I_y, \mathbf I_z)$ is the nuclear spin vector, and $\mathbf Q$ is the nuclear electric quadrupole tensor. This Hamiltonian can be simplified in the following way: First, we neglect the nuclear Zeeman energy (second term above) since its contribution is negligible at magnetic fields used in this work ($\simeq 10\,$mT). Second, the Jahn-Teller distortion defines a symmetry axis for the nitrogen defect along any of the [111]-crystal axis directions~\cite{Davies1981, Ammerlaan1981}. Under this trigonal symmetry (as with NV centers), and by going into an appropriate coordinate system, tensors $\mathbf g$, $\mathbf A$, and $\mathbf Q$ are diagonal and defined by at most two parameters:

\begin{equation}
  \mathbf g = \left( \begin{array}{ccc}
  g_\perp & 0 & 0 \\
  0 & g_\perp & 0 \\
  0 & 0 & g_\parallel
  \end{array} \right), \mathbf A = \left( \begin{array}{ccc}
  A_\perp & 0 & 0 \\
  0 & A_\perp & 0 \\
  0 & 0 & A_\parallel
  \end{array} \right) \text{, and } \mathbf Q = \left( \begin{array}{ccc}
  P_\perp & 0 & 0 \\
  0 & P_\perp & 0 \\
  0 & 0 & P_\parallel
  \end{array} \right).
\end{equation}

Here, $g_\perp$, $g_\parallel$, $A_\perp$, $A_\parallel$, $P_\perp$, and $P_\parallel$ are the gyromagnetic, hyperfine, and quadrupolar on- and off-axis tensor components, respectively, in the principal coordinate system. Further simplifications can be made by noting that the g-factor is approximately isotropic~\cite{Smith1959}, i.e., $g_\perp \approx g_\parallel \equiv g$, and that for exact axial symmetry the off-axis components of the quadrupole tensor, $P_\perp$, vanish~\cite{Slichter1990}. Equation\,\ref{eqn:p1hamil} may now be written as

\begin{equation} \label{eqn:p1hamil2}
H_\text{P1}/h = g \mu_B/h ~B_z ~\mathbf S_z + A_\parallel \mathbf S_z \cdot \mathbf I_z + A_\perp(\mathbf S_x \cdot \mathbf I_x + \mathbf S_y \cdot \mathbf I_y) + P_{||} (\mathbf I_z^2 - \mathbf{I}^3/3).
\end{equation}

\begin{figure}[ht]
  \centering	
  \includegraphics[width=0.75\textwidth]{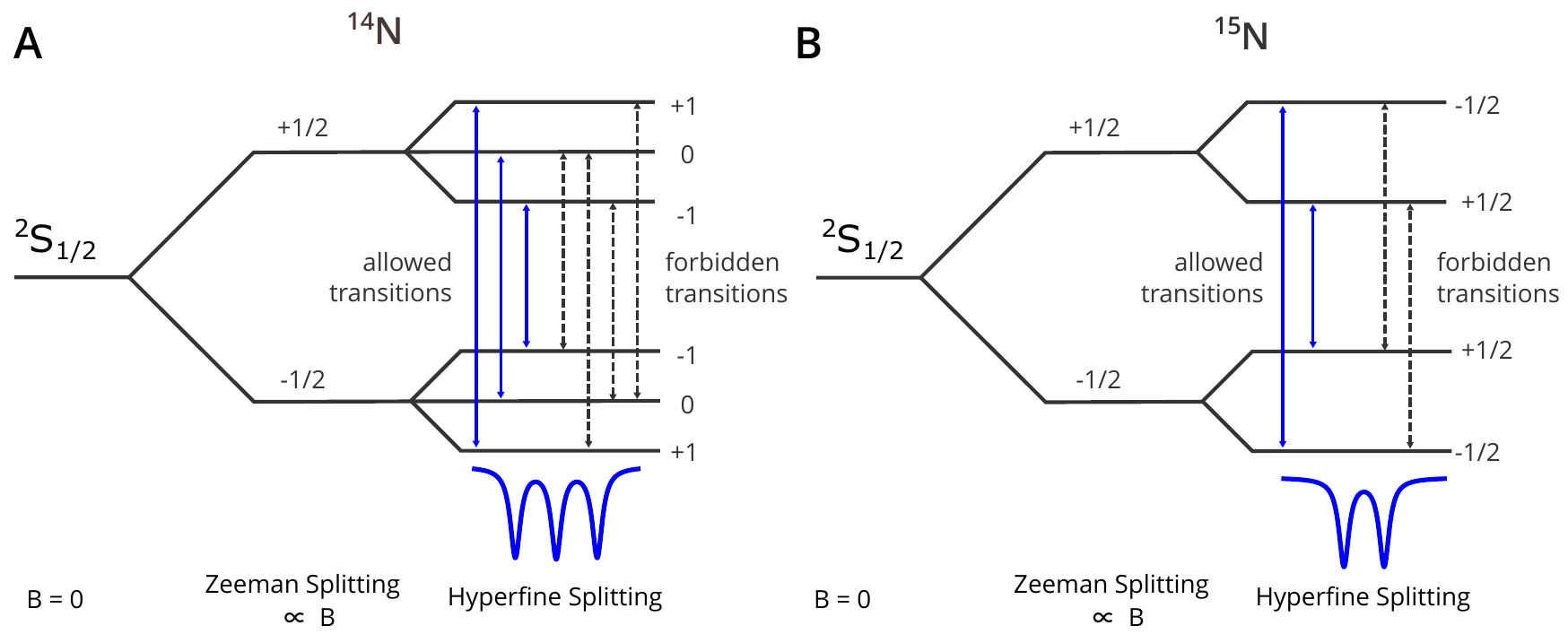}  
  \caption{
  {$^{14}$N and $^{15}$N spin energy level diagram.}
  }
  \label{fig:nleveldiagram}
\end{figure}

\subsection*{$^{14}$N spectrum}

$^{14}$N has $S=1/2$ and $I=1$, leading to six eigenstates $|m_S=\pm 1/2, m_I=0,\pm1 \rangle$. The corresponding three dipole-allowed transitions ($\Delta m_S=\pm 1, \Delta m_I=0$, solid arrows) are shown in Fig.\,\ref{fig:nleveldiagram}, along with the four first-order forbidden transitions ($\Delta m_S=\pm 1, \Delta m_I = \pm 1$, dashed arrows). A nitrogen defect in diamond undergoes a Jahn-Teller (JT) distortion, which defines a hyperfine quantization axis along any of the four [111] crystallographic directions, irrespective of the applied magnetic field. Taking all JT orientations into account, the full $^{14}$N spin resonance spectrum displays a total of 12 dipole-allowed resonances. By aligning the magnetic field along any of the [111]-directions of the diamond crystal, the 12 transitions are partially degenerate and reduce to six visible transitions in an NV DEER measurement, with an amplitude ratio 1:3:1:3:3:1, as shown in Fig.\,2b of the main text and Fig.\,\ref{fig:suppldeern14andn15}a. 
We obtain the spectrum for the off-axis and degenerate JT orientations from Eqn.\,\ref{eqn:p1hamil2} by rotating the bias field by $\theta = 109.471 $ around either the x or y axis, where $\theta$ is the angle between any two crystallographic axes, i.e., taking $\mathbf B \rightarrow \mathbf R_\text{x or y} (\theta = 109.471 \degree) \cdot \mathbf B.$\\

In Fig.\,\ref{fig:suppldeern14andn15}a, the simulated (degenerate) spectrum for $^{14}$N spins is shown together with experimental data from Sample B ($[\text N] = 0.75\,$ppm). A magnetic field $B_z = 95.5\,$mT is applied along one of the [111]-orientations. For the simulation the following parameters have been used: $g \mu_B/h \approx 2.8025 \times 10^4$ MHz/T, where $g =2.0025$ is the P1 electronic g-factor~\cite{Smith1959}, $\mu_B = 9.274 \times 10^{-24}$ J/T is the Bohr magneton, $h = 6.626 \times 10^{-34}$ Js is Planck's constant, $A_\parallel = 114$ MHz, $A_\perp = 81.3$ MHz~\cite{Cox1994,Cook1966,Smith1959},
and $P_{||} = -3.97\,$MHz~\cite{Cook1966}.

\begin{table}[ht]
\begin{center}
 \begin{tabular}{*2c} 
\hline
\hline \\ [-1.5ex]
\multicolumn{2}{c}{$^{14}$N} \\ 
\hline
g & 2.0025~\cite{Smith1959}\\
$A_\perp, A_{||}$ & 114\,MHz, 81.3\,MHz~\cite{Cox1994,Cook1966,Smith1959}\\
$P_{||}$ & -3.97\,MHz~\cite{Cook1966}\vspace{0.5 em}\\ 
\multicolumn{2}{c}{$^{15}$N} \\
\hline
$A_\parallel$, $A_\perp$ & -159.7\,MHz, -113.83\,MHz~\cite{Cox1994}\\
P & 0 (since $I < 1$)\\
\hline
\hline
\end{tabular}
\end{center}
\caption{
Summary of defect parameters used to simulate the nitrogen resonance spectrum using Eqn.\,\ref{eqn:p1hamil2}.
}
\label{tab:supplnitrogen}
\end{table}

\subsection*{$^{15}$N spectrum}

$^{15}$N has $S=1/2$ and $I=1/2$, leading to the four eigenstates $|m_S=\pm 1/2, m_I=\pm1/2 \rangle$. The corresponding two dipole-allowed transitions ($\Delta m_S=\pm 1, \Delta m_I=0$, solid arrows) are shown in Fig.\,\ref{fig:nleveldiagram}b, along with the two first-order forbidden transitions ($\Delta m_S=\pm 1, \Delta m_I = \pm 1$, dashed arrows). The experimental NV DEER spectrum for Sample C ([$^{15}$N]=10\,ppm) is shown in Fig.\,\ref{fig:suppldeern14andn15}b, along with a simulated $^{15}$N spectrum. For the $^{15}$N simulation we used $B_0=9.8\,$mT, $A_\parallel = -159.7$ MHz,  $A_\perp = -113.83$ MHz~\cite{Cox1994}, and $P_{||}=0$ (since $I < 1$).


\section{Continuous versus pulsed spin bath driving}\label{sec:pulsedvscwdriving}

\begin{figure}[ht]
  \centering
  \includegraphics[width=0.75\textwidth]{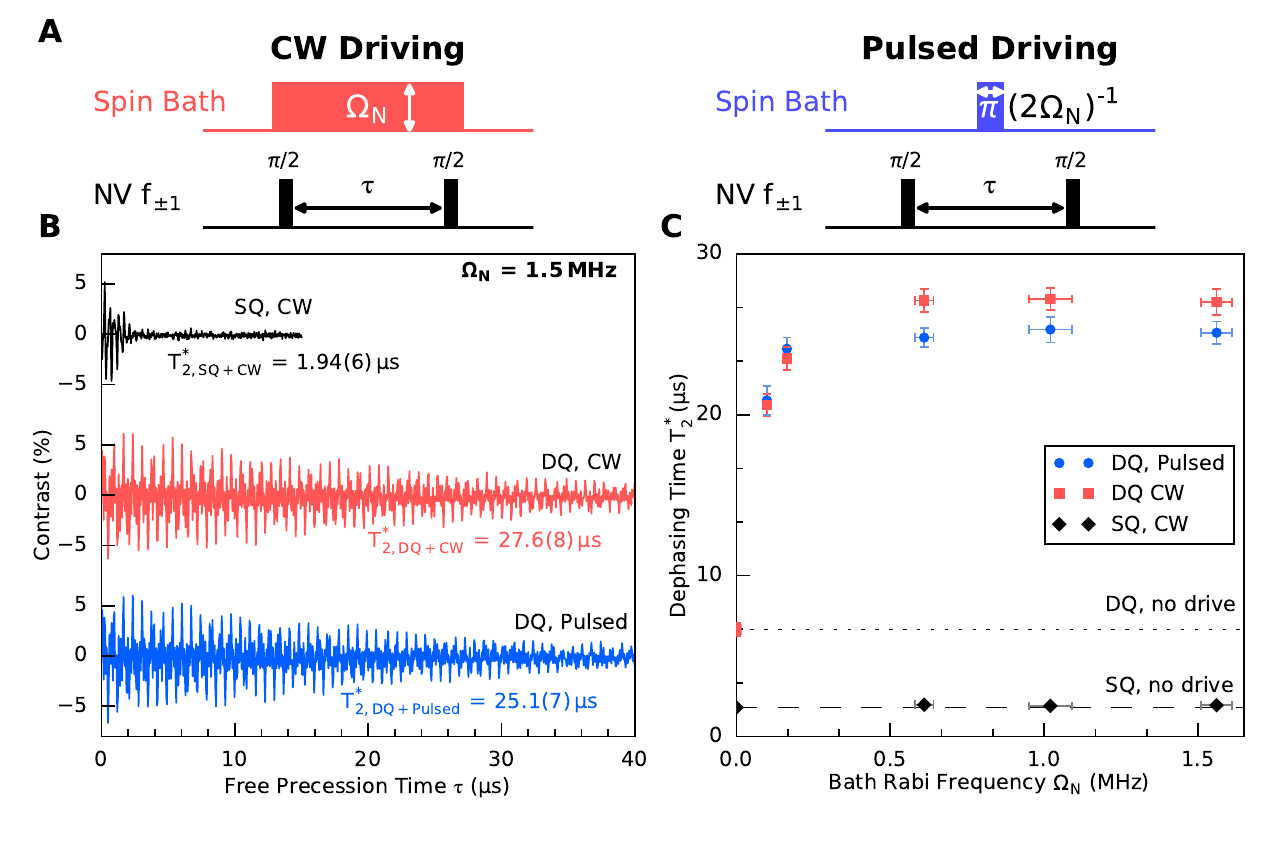}  
  \caption{{Comparison of CW to pulsed spin bath driving experiments using Sample B.} 
  {(a)} Sequences used for CW (left) and pulsed (right) decoupling of the electronic nitrogen spin bath. For both methods, six distinct frequencies are used to resonantly address the nitrogen spin bath via an applied RF field with equal Rabi frequency $\Omega_\text{N}$ on each spin transition. In {(b)} the Ramsey decay in the DQ basis for CW and pulsed driving are compared ($\Omega_\text{N}=1.5\,$MHz). The decay for CW driving in the SQ basis is included for reference. {(c)} Depicts $T_2^*$ as a function of bath Rabi frequency for DQ CW (red squares) and DQ pulsed (blue circles) spin bath driving, with the SQ CW results (black diamonds) again included for reference. The finely (coarsely) dashed line indicates the $T_2^*$ value in the DQ (SQ) basis without any drive field applied to the bath spins. 
  }
  \label{fig:SupplPulsedDriving}
\end{figure}

As described in the main text, both continuous (CW) and pulsed driving can decouple the electronic spin bath from the NV sensor spins (see Fig.\,\ref{fig:SupplPulsedDriving}). In CW driving, the bath spins are driven continuously such that they undergo many Rabi oscillations during the characteristic interaction time $1/\gamma_\text{NV-N}$, and thus the time-averaged NV-N dipolar interaction approaches zero. For pulsed driving, $\pi$-pulses resonant with spin transitions in the bath are applied midway through the NV Ramsey free precession interval, to refocus bath-induced dephasing. Fig.\,\ref{fig:SupplPulsedDriving}a illustrates both methods for a given applied RF field with a Rabi frequency of $\Omega_\text{N}$.\\ 

Although we treat CW driving in the main text in detail, we find experimentally that pulsed driving yields similar $T_2^*$ improvements over the measured range of Rabi drives. For example, Fig.\,\ref{fig:SupplPulsedDriving}a compares $T_2^*$ for Sample B for both schemes at maximum  bath drive strength $\Omega_\text{N} = 1.5\,$MHz (for pulsed driving $\tau_\pi \equiv 1/2\Omega_\text{N}$). Both decoupling schemes result in comparable $T_2^*$ improvements ($13-15\times$) over the non-driven SQ measurement, which is shown for reference. We attribute the slightly lower max $T_2^*$ achieved in pulsed driving to detunings of the RF drive from the spin resonances of the main nitrogen groups, leading to less efficient driving of the spin population (see next section). To study the efficacy of both driving schemes, we plot $T_2^*$ as a function of Rabi drive $\Omega_\text{N}$ in Fig.\,\ref{fig:SupplPulsedDriving}b. In the limit of $\tau_\pi \approx T_2^*$, pulsed driving resembles the CW case and both schemes converge to the same maximal $T_2^*$.\\

Despite the similar improvements in $T_2^*$ achieved using both methods, pulsed driving can reduce heating of the MW delivery loop and diamond sample - an important consideration for temperature sensitive applications. For this reason, pulsed driving may be preferable in such experiments despite the need for $\pi$-pulse calibration across multiple resonances.\\


\section{NV and nitrogen spin resonance linewidth measurements}\label{sec:nvp1linewidth}

\begin{figure}[ht!]
  \centering
  \includegraphics[width=0.65\textwidth]{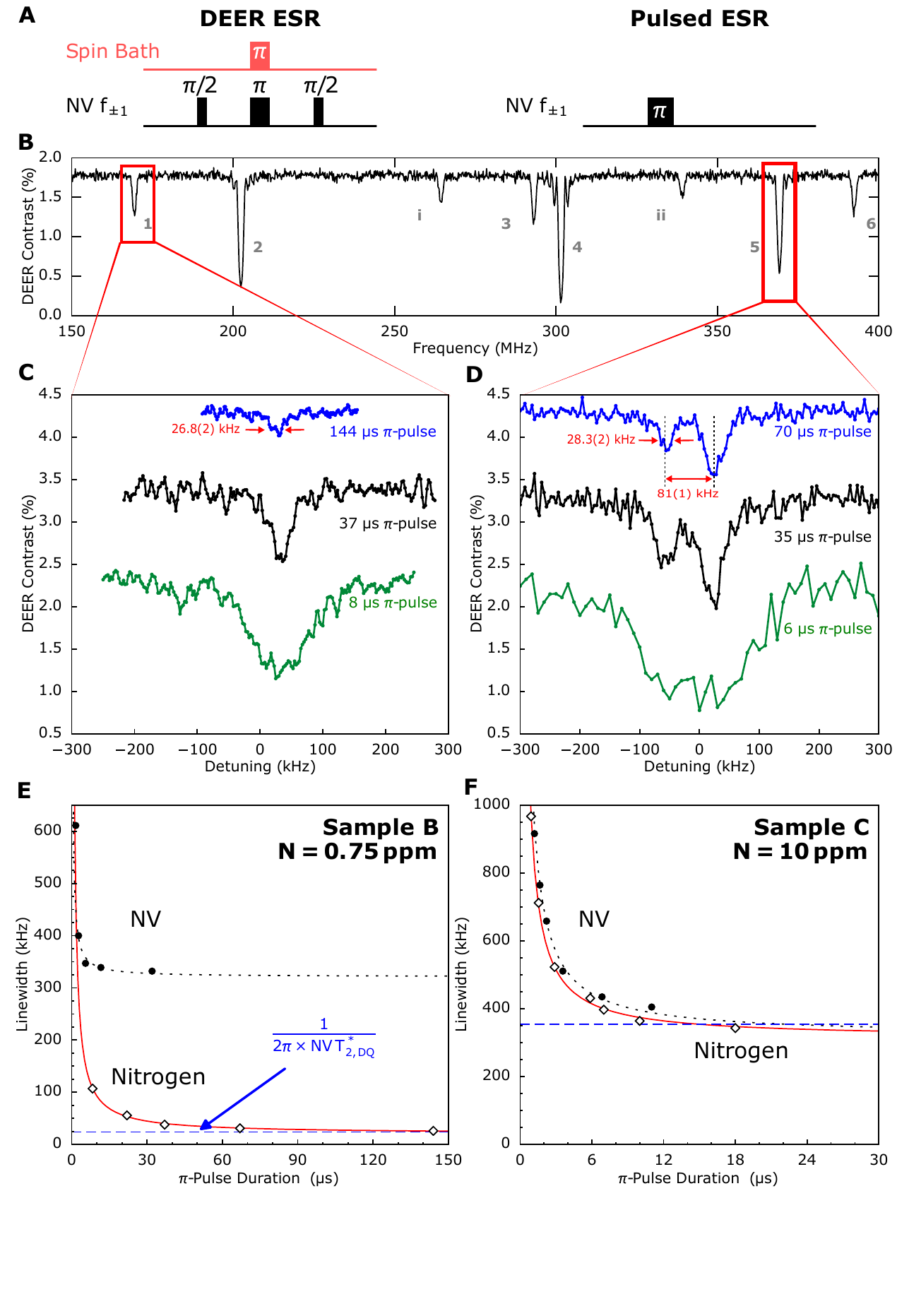}  
  \caption{{Comparison of nitrogen and NV spin resonance linewidths}. {(a)} The pulsed DEER (left) sequence used for spin resonance measurements of the nitrogen spins and the pulsed ESR (right) sequence is used for NV SQ spin resonance measurements.
  {(b)} DEER spectrum from 150 to 400 MHz including all six nitrogen transitions and two forbidden transitions in Sample B. {(c)} DEER spectra of a single nitrogen transition are shown for three different bath $\pi$-pulse durations. A minimum measured linewidth of 26.8(2) kHz was recorded using a $144 \, \upmu$s $\pi$-pulse. {(d)} DEER spectra for a group containing three nearly degenerate off-axis nitrogen transitions. When bath $\pi$-pulses of $70\,\upmu$s and $35\,\upmu$s are used, two features are resolved corresponding to a single nitrogen transition detuned by $81\,$kHz from two nearly overlapped transitions. {(e)} Comparison of the NV ESR linewidth (black dots) and the DEER linewidth for a single nitrogen transition (diamonds) as a function of $\pi$-pulse duration for Sample B ($[\text N]=0.75\,$ppm). The fine, black dashed and red solid lines correspond to fits of the NV and nitrogen spin resonance linewidths to the functional form $a/x+b$, where b is the saturation linewidth. The coarse, blue dashed line indicates the expected linewidth from the measured NV $T_2^*$ in the DQ basis (assuming a Lorentzian linewidth). {(f)} Same as (e) but for Sample C ($[\text N]=10\,$ppm).
  }
  \label{fig:overlapped}
\end{figure}

The NV and nitrogen (P1) ensemble spin resonance linewidths are determined using pulsed ESR and pulsed DEER NV spectral measurements, respectively, as shown in Fig.\,\ref{fig:overlapped}. Low Rabi drive strength and consequently long $\pi$-pulse durations can be used to avoid Fourier power broadening~\cite{Dreau2011}. We find that nitrogen spin resonance spectra are typically narrower than for NV ensembles in the SQ basis, due to the effects of strain gradients in diamond on NV zero-field splittings.\\ 

For the spin bath driving model described in the main text (Eqn. 4), we are interested in the natural (i.e., non-power-boadened) linewidth $\delta_\text{N}$ of spin resonances corresponding to, for example, $^{14}$N groups $1-6$ (see Fig.\,2b 
in main text and Fig.\,\ref{fig:overlapped}a in Supplement). In Ref.~\cite{DeLange2010} it was reported that the different $^{14}$N groups have approximately equal linewidth, i.e., that $\delta_{\text{N},i} \approx \delta_\text{N}$. However, we find that the bias field $B_z$ being only slightly misaligned ($\sim$3 degree) from one of the [111] crystal axes causes the three degenerate spin resonances to be imperfectly overlapped, leading to a larger effective linewidth.\\

In Fig.\,\ref{fig:overlapped}b and c we compare the NV pulsed DEER linewidths of $^{14}$N  group 1 (a single resonance) with that of group 5 (three overlapped resonances) for different $\pi_\text{bath}$-pulse durations. At short $\pi_\text{bath}$-pulse durations (high MW powers), the linewidths are power broadened due to the applied microwave field, such that the measured linewidth is a convolution of the natural linewidth and the inverse duration of the $\pi_\text{bath}$-pulse~\cite{Dreau2011}. At longer $\pi_\text{bath}$-pulse durations (reduced MW power), however, the measured linewidth approaches its natural width. In this instance, and for dipolar-limited linewidth broadening, the lineshape is Lorentzian with full width at half max $\Gamma = 1/\pi T_{2,\text{N}}^{*}$. At the longest $\pi_\text{bath}$-pulse durations used in this work, we find that group 1 consists of a narrow, approximately 25\,kHz-wide peak. In contrast, group 5 reveals two peaks, consisting of two overlapped $^{14}$N transitions and one detuned transition, which is attributed to imperfect magnetic field alignment. The splitting between the two peaks in group 5 is $\approx 80\,$kHz, which we use as the effective $^{14}$N linewidth $\delta_\text{N}$ in Eqn.~4 of the main text, and which is consistent with the value extracted from fitting the spin-bath driving model to the data (see Fig.\,4a, 
$\delta_\text{N}^\text{fit} \approx 60\,$kHz).\\

In Fig.\,\ref{fig:overlapped}e we compare the measured NV and $^{14}$N group 1 ensemble linewidths (full width at half max) for Sample B as a function of $\pi$-pulse duration. For both species, the linewidth narrows at long $\pi$-pulse durations, as discussed above, reaching non-power-broadened (natural) values. Notably, the non-power-broadened NV linewidth [$321(7)\,$kHz, extracted from a fit to the data] is $\sim 16\times$ larger than the natural $^{14}$N linewidth [$20.6(1.2)\,$kHz]. This order-of-magnitude difference is a manifestation of the strong strain field gradients in this sample. Specifically, pulsed ESR measurements of the NV ensemble linewidth (see Fig.\,\ref{fig:overlapped}a) are performed in the SQ $\{0,+1\}$ or $\{0,-1\}$ sub-basis, and are therefore strain gradient limited. In contrast, nitrogen defects in diamond have $S=1/2$, and thus do not couple to electric fields or strain gradients. As a consistency check, note that NV ensemble Ramsey measurements in Sample B, made in the DQ basis (with no spin-bath driving), yield a strain-independent dephasing time $T_{2,\text{DQ}}^*=6.9(5)\,\mu$s. This dephasing time, presumably limited by the nitrogen spin bath, implies a $^{14}$N spin resonance linewidth given by $\frac{1}{2} \times 1/\pi T_{2,\text{DQ}}^* = 23(2)\,$kHz, which is in good agreement with our pulsed DEER measurements of the natural $^{14}$N linewidth. Similar consistency is found for measurements of the NV and $^{15}$N ensemble spin resonance linewidths in Sample C, as shown in Fig.\,\ref{fig:overlapped}f. Such agreement across multiple samples is further evidence that the DQ $T_2^*$ value for NV ensembles is limited by the surrounding nitrogen spin bath, as discussed in the main text. Note that for our samples $[\text{NV}] \ll [\text N]$ and we can therefore ignore the back action of NVs onto nitrogen spins in the DEER readout. For denser NV samples, however, this back action has to be taken into account~\cite{Stepanov2016}. \\

\section{DC magnetometry with DQ and spin-bath drive}\label{sec:dcsense}

Assuming a signal-to-noise ratio of unity, the minimum detectable magnetic field $\delta B_{min}$ in a Ramsey measurement is given by~\cite{Fang2013}
\begin{equation}\label{eqn:expsensitivity}
	\delta B_{min} \approx \frac{\delta S}{\max | \frac{\partial S}{\partial B} |},
\end{equation}
where the Ramsey signal $S$ is
\begin{equation}
	S = C(\tau) \sin(\gamma_\text{NV} B_{DC} \tau).
\end{equation}

Here, {$C(\tau) = C_0 \exp{(-(\tau / T_2^*)^p)}$} is the time-dependent measurement contrast defined via the NV spin-state-dependent fluorescence visibility (see Suppl.\,\ref{sec:expmethods}), $\gamma_\text{NV}$ is the NV gyromagnetic ratio, $B_{DC}$ is the magnetic field to be sensed, and $\tau$ is the sensing time during which the NV sensor spins accumulate phase. The term $\max | \frac{\partial S}{\partial B} |$ is the maximum slope of the Ramsey signal,
\begin{equation}
\max | \frac{\partial S}{\partial B} | = C(\tau) \gamma_\text{NV} \tau. 
\end{equation}

Assuming uncorrelated, Gaussian noise, $\delta S = \sigma(\tau)/\sqrt{n_{meas}}$ is the standard error of the contrast signal, which improves with number of measurements $n_{meas}$. Including a dead time $\tau_D$ that accounts for time spent during initialization of the NV ensemble and readout of the spin-state-dependent fluorescence during a single measurement, $n_{meas} = T/(\tau + \tau_D)$ measurements are made over the total measurement time $T$. $\delta  B_{min}$ is then found to be 
\begin{equation}
	\delta B_{min} = \frac{\sigma \sqrt{\tau + \tau_D}}{C(\tau) \gamma_\text{NV} \tau \sqrt{T}}, 
\end{equation}
and the sensitivity is given by multiplying $\delta  B_{min}$ by the bandwidth $\sqrt{T}$ and including a factor $\Delta m$ = 1(2) for the SQ (DQ) basis: 
\begin{equation} \label{eqn:ramseysensitivity2}
	\eta =  \frac{\delta B_{min} \sqrt{T}}{\Delta m} = \frac{\sigma \sqrt{\tau + \tau_D}}{\Delta m \times C(\tau) \gamma_\text{NV} \tau}.
\end{equation}
Note that in the ideal case, $\tau_{D} \ll \tau$, we have $\frac{\sqrt{\tau +\tau_D}}{\tau} \approx 1/\sqrt{\tau}$ and the sensitivity $\eta$ scales {$\propto \tau^{-1/2} \exp{(\tau / T_2^*)^p}$}. The optimal sensing time in our Ramsey experiment is then {$\tau_{opt} \approx T_2^*/2$ for $p = 1-2$}. However, in the more realistic case, $\tau_D \sim \tau$, the improvement of $\eta$ with increasing $\tau$ approaches a linear scaling and {$\eta \propto \tau^{-1} \exp{(\tau / T_2^*)^p}$ for $\tau_D \gg \tau$}. The optimal sensing time then becomes $\tau_{opt} \approx T_2^*$. Consequently, the measured increase in sensitivity may exceed the enhancement estimated from the idealized case without overhead time.\\

\begin{figure}[ht]
  \centering
  \includegraphics[width=0.75\textwidth]{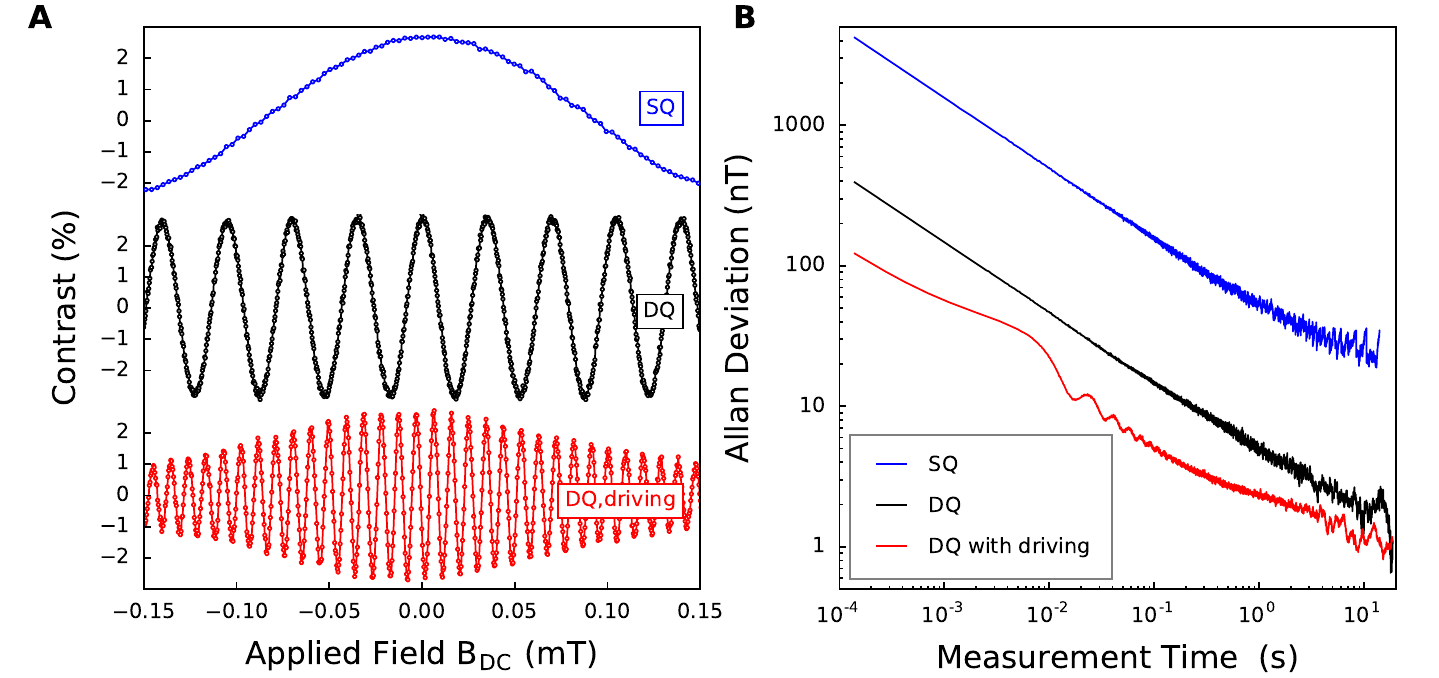}  
  \caption{{DC magnetic field sensing and Allan deviation.} {(a)} DC magnetometry curves for SQ, DQ, and DQ with spin-bath driving in Sample B, produced by sweeping the magnitude of a coil-generated applied magnetic field (in addition to the fixed bias field) while the free precession interval $\tau$ is set to $\tau_{SQ} = 1.308\, \upmu$s (blue, top), 
  $\tau_{DQ} = 6.436\, \upmu$s (black, middle), and $\tau_{DQ,drive} = 23.990\, \upmu s$ (red, bottom). Reproduced from Fig.\,4d
{(b)} Allan deviation using the same fixed $\tau$ values from (a) for measurements using SQ (blue), DQ (black), and DQ with driving (red). The external field strength was tuned to sit on a zero crossing of the respective DC magnetometry curves in (a) for sine magnetometry. 
  }
  \label{fig:DCsense}
\end{figure}

With Eqn.\,\ref{eqn:ramseysensitivity2} we calculate and compare the sensitivities for the three measurement modalities (SQ, DQ, and DQ + spin-bath drive) applied to Sample B. Using $C\approx 0.026$, which remains constant for the three schemes (see Fig.\,\ref{fig:DCsense}a), sensing times $\tau_{SQ} = 1.308\,\upmu$s, $\tau_{DQ} = 6.436\,\upmu$s, and $\tau_{DQ+Drive} = 23.99\,\upmu$s, standard deviations $\sigma_{SQ}=0.0321$, $\sigma_{DQ} =0.0324$, and $\sigma_{DQ+Drive} = 0.0325$ calculated from $1\,$s of data, fixed sequence duration of $\tau+\tau_{D} = 70\,\upmu$s, and $\gamma_\text{NV} = 2\pi \times 28\,$GHz/T, the estimated sensitivities for the SQ, DQ and DQ+Drive measurement schemes are $\eta= 70.7$, $6.65$, and $1.97\,\mathrm{nT}/\sqrt{\mathrm{Hz}}$, respectively. In summary, we obtain a $10\times$ improvement in DC magnetic field sensitivity in the DQ basis, relative to the conventional SQ basis, and a $35\times$ improvement using the DQ basis with spin bath drive. Note that this enhancement greatly exceeds the expected improvement when no dead time is present ($\tau_D \ll \tau$) and is attributed to the approximately linear increase in sensitivity with {sensing times $\tau \lesssim \tau_D$}. We also plot the Allan deviation for the three schemes in Fig.\,\ref{fig:DCsense}b showing a $\tau^{-1/2}$ scaling for a measurement time of $\approx 1\,$s and the indicated enhancements in sensitivity.
%
%

{\
Lastly, we discuss an appropriate choice of spin concentrations in diamond and other sample material properties for enhanced-sensitivity magnetometry employing DQ coherence and spin bath driving. To simplify the discussion, we focus on the following combination of relevant parameters for NV magnetometry, which (in the appropriate limits discussed herein) is proportional to the photon-shot-noise-limited volume-normalized magnetic sensitivity $\eta^\text{V}$ (see Suppl. of Ref.~\cite{Barry2016}), given by
\begin{equation}\label{eqn:suppl_fom}
	\eta^\text{V} \propto  \left[ \Delta m \times \sqrt{n_\text{NV} \cdot [\mathrm N] \cdot T_2^*([\mathrm N)])} \right]^{-1} \equiv \eta_\text{N}.
\end{equation}
Here, $\Delta m = 1(2)$ in the SQ (DQ) basis, [N] is the substitutional nitrogen (P1) center concentration, $n_\text{NV} = $[\text{NV}]/[\text{N}] is the normalized concentration of NV centers relative to the nitrogen concentration, and $T_{2}^*([\mathrm N])$ is the NV ensemble dephasing time in the sensing basis chosen (SQ or DQ). The quantity $\eta_\text{N}$ describes the dependence of the sensitivity on nitrogen concentration $[\mathrm N]$. Since the nitrogen concentration enters Eqn.\,\ref{eqn:suppl_fom} both explicitly and also through $n_\text{NV}$ and $T_2^*$, we need to investigate Eqn.\,\ref{eqn:suppl_fom} for a range of [N].

In the case of fixed $n_\text{NV}$ and nitrogen-spin-bath-limited dephasing, i.e., $T_2^* \propto 1/[\mathrm N]$, $\eta_\text{N}$ remains constant as a function of [N]. In this simplistic picture, shorter $T_2^*$ values may be exchanged for higher nitrogen (and thus NV center) concentrations and vice versa, with no effect on sensitivity. Such a discussion, however, neglects experimental overhead due to NV state initialization and readout, which is characterized by the dead time $\tau_{D}$ (see Eqn.\,\ref{eqn:ramseysensitivity2}). Since $T_2^* \lesssim 1\,\upmu\mathrm{s} \lesssim \tau_D$ in a typical SQ NV ensemble experiment, increasing $T_2^*$ through optimized sample fabrication, DQ coherence magnetometry, and/or spin bath driving is preferred, and larger sensitivity gains are obtained when compared to an equivalent increase in NV center concentration.

More generally, the ensemble $T_2^*$ depends on numerous diamond-related parameters (including the concentration of spin impurities and strain fields) and external conditions (such as temperature fluctuations of the diamond sample and magnetic field gradients due to the applied bias field). Focusing on the parameters intrinsic to diamond, the relevant contributions to $T_2^*$ are (compare to Eqn.\,1 in main text)
\begin{equation}\label{eqn:suppl_fom2}
1/T_2^* \approx 1/T_2^*\{\text{NV-}^{13}\mathrm C\} + 1/T_2^*\{\text{NV-N}\}(\Omega_\text{N}) + 1/T_2^*\{\text{NV-NV}\}(\mathrm N) + 1/T_2^*\{\text{strain}\} + ...,
\end{equation}
where we added the term $1/T_2^*\{\text{NV-NV}\}$ to account for dephasing due to NV-NV dipolar interactions. This dephasing mechanism was neglected in the main text due to the low N-to-NV conversion efficiencies of Samples A, B, and C ($n_\text{NV}\ll 1$), but its contribution becomes relevant at increased conversion efficiencies intended for optimized diamond magnetometry. To model Eqn.\,\ref{eqn:suppl_fom} across a range of nitrogen concentrations, we now combine Eqns.\,\ref{eqn:suppl_fom} and \ref{eqn:suppl_fom2} and include the dependence of $T_2^*\{\text{NV-N}\}(\Omega_\text{N})$ on bath drive strength $\Omega_\text{N}$ (see Eqn.\,5 in the main text). We also anticipate optimized diamond samples to be isotopically engineered with $T_2^*\{^{13}\mathrm C = 0.01\,\% \} \simeq 100\,\upmu$s (or longer) and to possess strain field gradients comparable to this work's samples ($T_2^*\{\text{strain}\} \simeq 5\,\upmu$s). N-to-NV conversion efficiencies of up to $30\,\%$ have been reported for NV ensembles~\cite{Wolf2015} suggesting that $n_\text{NV} \simeq 0.4$ is feasible for an optimized diamond sample. The simulation results for $\eta_\text{N}$ in this parameter regime are summarized in Fig.\,\ref{fig:sensitivity_comparison} for SQ (blue) and DQ coherence magnetometry (red) and plotted for spin bath drive strengths $\Omega_\text{N}$ = 0 (solid), 1, and 10\,MHz (dashed).

\begin{figure}[ht]
  \centering
  \includegraphics[width=0.75\textwidth]{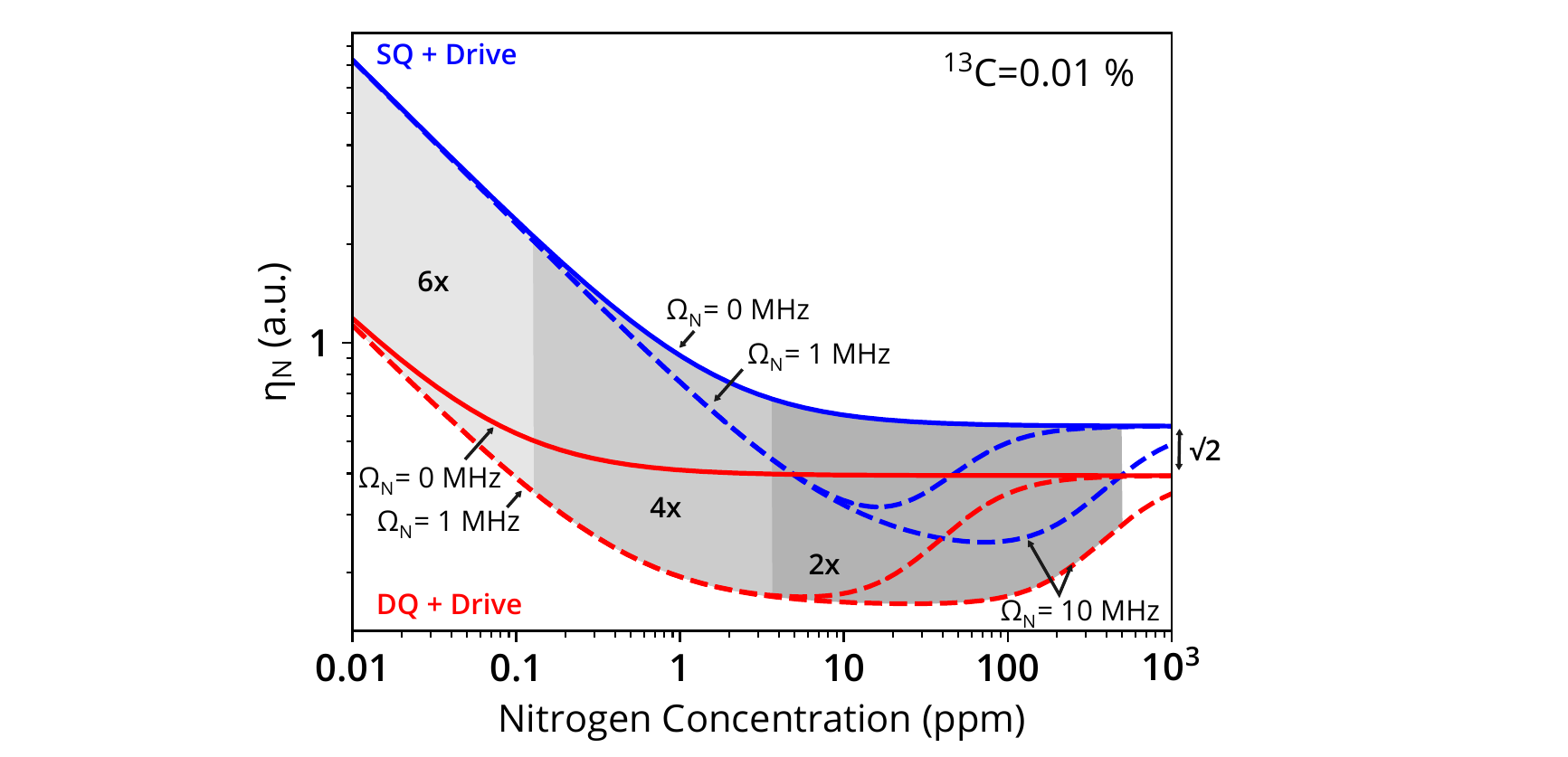}  
  \caption{ \
  $\eta_\text{N}$ given by Eqn.\,\ref{eqn:suppl_fom}, which is proportional to the NV ensemble volume-normalized magnetic field 
  sensitivity, as a function of nitrogen (P1) center concentration [N] for SQ (blue) and DQ (red) magnetometry with spin bath drive strengths $\Omega_\text{N} = 0$ (solid), 1, and 10\,MHz (dashed). Lower values correspond to higher sensitivities and vice versa. For the simulation we combine Eqns.\,\ref{eqn:suppl_fom} and \ref{eqn:suppl_fom2} with the following parameters: $T_2^*\{^{13}\mathrm C = 0.01\,\% \} \simeq 100\,\upmu$s, $T_2^*\{\mathrm{strain}\} \simeq 5\,\upmu$s, $T_2^*\{\text{NV-N}\}$ is given by Eqn.\,5 of the main text, $n_\text{NV} = 0.4$, and $T_2^*\{\text{NV-NV}\}(\mathrm N) = (A_\text{NV-NV} \cdot n_\text{NV} \cdot [\mathrm N]/4)^{-1}$. Here, $A_\text{NV-NV} \approx 2 A_\text{NV-N} \simeq 2\pi \times 33\,$kHz/ppm due to the twice higher spin multiplicity of the NV centers \cite[Ch. III and IV]{Abragam1983}, and the factor 1/4 accounts for the fraction of NV centers used for sensing when all four NV orientations are distinguished. In this instance and assuming perfect optical initialization of NV centers, 3/4 of the NV centers are in the $m_s=0$ spin state and do not contribute to dephasing during the sensing sequence. Grey shaded regions indicate approximate improvements in sensitivity for DQ magnetometry with drive over SQ magnetometry alone.
  }
  \label{fig:sensitivity_comparison}
\end{figure}

Without spin bath drive applied ($\Omega_\text{N} = 0$) and at low nitrogen concentrations ($[\text N]\lesssim 1\,$ppm), $\eta_\text{N}$ in Eqn.\,\ref{eqn:suppl_fom} is larger (i.e., sensitivity is reduced) for SQ magnetometry due to the $T_2^*$-limit imposed by strain gradients. In this lower nitrogen regime, working in the DQ basis leads to substantial improvements in sensitivity (i.e., smaller values and thus higher sensitivity). At higher nitrogen concentrations ($[\text N] \gtrsim 1\,$ppm), dipolar interactions with the spin bath dominate NV dephasing, strain contributions become negligible, and DQ coherence approaches a $\sqrt{2}$ enhancement in sensitivity over SQ coherence measurements. Note that the crossover between the low and high nitrogen regime is set by the strain contributions to $T_2^*$ (here $\simeq 5\,\upmu$s) and lower (higher) strain contributions  
shift the crossover to lower (higher) nitrogen concentrations.

With spin bath drive applied, however, we find that additional gains in sensitivity are obtained for both SQ and DQ coherence measurements. The largest improvements for $\eta_\text{N}$ are obtained when DQ coherence measurements are employed at high spin bath drive strengths ($1-10\,$MHz) and for an optimal nitrogen regime of $1 - 100\,$ppm. In practice, we expect the $1 - 10$\,ppm nitrogen regime to be optimal when all parameters relevant for magnetic sensing are considered. Herein we discuss additional parameters only qualitatively: i) The necessary Rabi frequency for effective spin bath driving increases linearly with nitrogen concentration, meaning the required RF power increases quadratically. Spin bath driving at high nitrogen concentrations becomes thus increasingly more challenging. ii) Samples A, B, and C in the main text were selected to have a predominately electronic nitrogen (P1) spin bath. The incorporation of high nitrogen concentrations in diamond, however, can lead to a larger variety of nitrogen-related spin species in the bath, which include nitrogen-clusters, NV$^0$, and NVH defects (for example see Refs. \cite{Zaitsev2001,Beha2012,Glover2004} and therein). A more diverse spin bath severely increases the complexity of the bath drive, eventually rendering it impractical. Finally, iii) there are indications that diamond samples with a high nitrogen content exhibit a larger fraction of NV$^0$ centers. In such samples, the NV$^-$ measurement contrast $C$ (see Eqn.\,\ref{eqn:ramseysensitivity2}) and, thus magnetic sensitivity, is diminished. 

Summarizing, our analysis suggests that the $1 - 10\,$ppm nitrogen regime is optimal for high sensitivity magnetometry using NV ensembles but further work is required to quantitatively account for all parameters relevant for sensing. Note that the $2\times$, $4\times$, and $6\times$ enhancement in sensitivity indicated in Fig.\,\ref{fig:sensitivity_comparison} for DQ with drive over SQ magnetometry alone, corresponds to a $4\times$, $16\times$, and $36\times$ reduction in measurement time, respectively; and even larger relative enhancements in sensitivity should be realized when accounting for the experimental dead time $\tau_D$ in NV ensemble experiments.
}

\section{Dephasing channels per sample}

\begin{table}[h]
  \centering
  \caption{
  {NV spin ensemble dephasing mechanisms for Sample A.}
  Individual contributions to dephasing are determined using the estimated/calibrated values described in the main text and Supplement (column 2). The data show good agreement between calculated and measured total dephasing times $T_{2,\text{SQ}}^*$ and $T_{2,\text{DQ}}^*$ (last two rows). 
  }
    \begin{tabular}{c c c c c}
 \hline
\hline
        Channel  & Magnitude & \multicolumn{2}{c}{Dephasing}       & Method \\
          &       & 1/$\upmu$s  & $\upmu$s    & \\
\hline \\[-1.5ex]
    strain &   0.0028 MHz/$\upmu$m
    & 0.190 & 5     & estimate \\
    $^{14}$N & 0.05 ppm &  0.0029  & 348   & dipolar estimate \\
    $^{13}$C & 0.01\% & 0.01  & 100   & calibration \\
    magnetic field gradient @ 20 G & 0.000056 MHz/G & 0.00112 & 893   & estimate \\
    \hline
    total SQ & & 0.2035 & 4.9 & $5-12\,\upmu$s (measured)\\
    total DQ $\times 2$ (no strain) & & 0.014 & 71 & 68 $\upmu$s (measured)\\
\hline
\hline
    \end{tabular}%
  \label{tab:dephasingsamplea}%
\end{table}%

\begin{table}[h]
  \centering
  \caption{
  {NV spin ensemble dephasing mechanisms for Sample B.}
  Similar to Sample A. Additionally, spin echo double-electron resonance (SEDOR) measurements were performed to estimate dephasing contributions from individual nitrogen resonance lines (for details see Ref.~\cite{DeLange2012}).
  }
    \begin{tabular}{c c c c c}
\hline
\hline
        Channel  & Magnitude & \multicolumn{2}{c}{Dephasing}       & Method \\
          &       & 1/$\upmu$s  & $\upmu$s    & \\
\hline \\[-1.5ex]
    
    strain & 0.0028 MHz/$\upmu$m & 0.190 & 5     & estimate \\
    $^{14}$N (allowed) &  & 0.056 & 18    &  SEDOR \\
    $^{14}$N (forbidden) &       & 0.0047 & 214   & SEDOR \\
    $^{13}$C   & 0.01\% & 0.01  & 100   & calibration \\
    magnetic field gradient @ 85 G & 0.000056 MHz/G & 0.00474 & 210   & estimate \\

\hline
    total SQ & & 0.265 & 3.8 & $1-10\,\upmu$s (measured)\\
    total DQ $\times 2$ (no strain) & & 0.076 & 13.1 & 13.8 $\upmu$s (measured)\\
\hline
\hline
    \end{tabular}%
  \label{tab:dephasingsampleb}%
\end{table}%

\begin{table}[h]
  \centering
  \caption{
  NV spin ensemble dephasing mechanisms for Sample C, similar to Samples A and B.
  }
    \begin{tabular}{c c c c c}
\hline
\hline
        Channel  & Magnitude & \multicolumn{2}{c}{Dephasing}       & Method \\
          &       & 1/$\upmu$s  & $\upmu$s    & \\
\hline \\[-1.5ex]
    strain & 0.0028 MHz/$\upmu$m & 0.140 & 7     & estimate \\
    $^{15}$N (allowed) & \multirow{5}[0]{*}{} & 0.59  & 2    & SEDOR \\
    $^{15}$N (forbidden) &       & 0.15  & 7     & SEDOR \\
    $^{14}$N (5\% of N15) &       & 0.0391  & 26    & estimated   \\
    $^{13}$C   & 0.05\% & 0.05  & 20    & calibrated \\
    magnetic field gradient @ 100 G & 0.000022 MHz/G & 0.0022 & 446   & estimate \\
\hline
    total SQ & & 1.01 & 1.0 & $0.3 - 1.2\,\upmu$s (measured)\\
    total DQ $\times 2$ (no strain) & & 0.87 & 1.1 & 1.2 $\upmu$s (measured)\\
\hline
\hline
    \end{tabular}%
  \label{tab:dephasingsamplec}%
\end{table}%


